\documentclass[aps,prd,a4paper,eqsecnum,superscriptaddress,nofootinbib,
showpacs,showkeys,amsfonts,amssymb,amsmath]{revtex4}

\usepackage{amssymb,latexsym}
\usepackage{amsmath, amsthm}
\usepackage{makeidx}
\usepackage{amscd}
\usepackage{amsthm}                
\usepackage{times}
\usepackage{epsfig}
\usepackage{psfrag}
\usepackage{graphicx}
\usepackage{amssymb,latexsym}
\numberwithin{equation}{section}
\makeindex

\begin{document}
\title[Genericity of naked singularities]{Genericity aspects in
gravitational collapse to black holes and naked singularities}

\author{Pankaj S. Joshi} \email{psj@tifr.res.in}
\affiliation{Tata Institute of Fundamental Research, Homi Bhabha Road,
Colaba, Mumbai 400005, India}
\author{Daniele Malafarina} \email{daniele.malafarina@polimi.it}
\affiliation{Tata Institute of Fundamental Research, Homi Bhabha Road,
Colaba, Mumbai 400005, India}
\author{Ravindra V. Saraykar} \email{r_saraykar@rediffmail.com}
\affiliation{Department of Mathematics, R.T.M. Nagpur University,
University Campus, Nagpur 440033, India}

\swapnumbers

\begin{abstract}
We investigate here the genericity and stability aspects
for naked singularities and black holes that arise as the final
states for a complete gravitational collapse of a spherical massive
matter cloud. The form of the matter considered is a general
{\it Type I} matter field, which includes most of the physically
reasonable matter fields such as dust, perfect fluids and
such other physically interesting forms of matter widely used in
gravitation theory.
We first study here in some detail the effects of small
pressure perturbations in an otherwise pressure-free collapse
scenario, and examine how a collapse evolution that was going
to the black hole endstate would be modified and go to a
naked singularity, once small pressures are introduced in
the initial data. This allows us to understand the distribution
of black holes and naked singularities in the initial data space.
Collapse is examined in terms of the evolutions
allowed by Einstein equations, under suitable physical conditions
and as evolving from a regular initial data.
We then show that both black holes and
naked singularities are generic outcomes of a complete collapse,
when genericity is defined in a suitable sense in an appropriate
space.

\end{abstract}
\maketitle

\section{Introduction}

The aim of the present work is twofold. Firstly, we study
here the effect of introducing small pressure perturbations in
an otherwise pressure-free gravitational collapse which was
to terminate in a black hole final state. For such a purpose,
spherically symmetric models of black hole and naked singularity
formation for a general matter field are considered, which
undergo a complete gravitational collapse under reasonable
physical conditions while satisfying suitable energy conditions.
Secondly, we investigate the genericity and stability
aspects of the occurrence of naked singularities and black holes
as collapse endstates. The analysis of
pressure perturbations in known collapse models, inhomogeneous
but otherwise pressure-free, shows how collapse final states
in terms of black hole or naked singularity are affected and altered.
This allows us to examine in general how generic these outcomes
are and we study in the initial data space the set of conditions
that lead the collapse to a naked singularity and investigate
how `abundant' these are. While it is known now for some time
that both black holes and naked singularities do arise
as collapse endstates under reasonable physical conditions,
this helps us understand and analyze in a clear manner the
genericity aspects of occurrence of these objects in
a complete gravitational collapse of a massive matter cloud
in general relativity.

The physics that is accepted today as the backbone of the
general mechanism describing the formation of black holes as the
endstate of collapse relies on the very simple and widely studied
Oppenheimer-Snyder-Datt (OSD) dust model, which describes the
collapse of a spherical cloud of homogeneous dust
\cite{OSD}, \cite{Datt}.
In the OSD case, all matter falls into the singularity at the same
comoving time while an horizon forms earlier than the singularity,
thus covering it. A black hole results as the endstate of collapse.
Still, homogeneous dust is a highly idealized and unphysical model
of matter. Taking into account inhomogeneities in the initial density
profile it is possible to show that the behaviour of the horizon
can change drastically, thus leaving two different outcomes as the
possible result of generic dust collapse: the black hole, in which
the horizon forms at a time anteceding the singularity, and the
naked singularity, in which the horizon is delayed thus allowing
null geodesics to escape the central singularity where the density
and curvatures diverge, to reach faraway observers
\cite{dust1}-\cite{dust5}.

It is known now that naked singularities do arise
as a general feature in General Relativity under a wide variety
of circumstances. Many examples of singular spacetimes can be
found, but their relevance in models describing physically viable
scenarios has been a matter of much debate since the first
formulation of the Cosmic Censorship Hypothesis (CCH)
\cite{Penrose}.
In particular, the formation of naked singularities in dynamical
collapse solutions of Einstein field equations remains a much
discussed problem of contemporary relativity. The CCH, which states
that any singularity occurring in the universe must be hidden within
an event horizon and therefore not visible to faraway observers,
has remained at the stage of a conjecture for more than four
decades now. This is also because of the difficulties lying in
a concrete and definitive formulation of the conjecture itself.
While no proof or any mathematically rigorous formulation of
the same exists
in the context of dynamical gravitational collapse
(while some proofs exist for particular classes of spacetimes
that do not describe gravitational collapse, as in
\cite{Dafermos1} and \cite{Dafermos2}),
many counterexamples have been found over the
past couple of decades
\cite{Ref1}-\cite{Ref6}.
Many of these collapse scenarios are restricted by some
simplifying assumptions such as the absence of pressures (dust
models) or the presence of only tangential pressures
\cite{tang1}-\cite{tang7}.
It is well-known that the pressures cannot be neglected in
realistic models describing stars in equilibrium. It seems natural
therefore that if one wishes to study analytically what happens
during the last stages of the life of a massive star, when its
core collapses under its own gravity thus forming a compact object
as a remnant, pressures must be taken into account.

Therefore, further to early works that showed the occurrence
of naked singularities in dust collapse, much effort has been
devoted to understanding the role played by pressures
\cite{press1}-\cite{press7}.
The presence of pressures is a crucial element towards the description
of realistic sources as we know that stars and compact objects are
generally sustained by matter with strong stresses (either isotropic
or anisotropic). At first it was believed that the naked singularity
scenario could be removed by the introduction of pressures, thus implying
that more realistic matter models would lead only to the formation of a
black hole. We now know that this is not the case. The final outcome
of collapse with pressure is entirely decided by its initial
configuration and allowed dynamical evolutions and it
can be either a black hole or a naked singularity. Furthermore it
is now clear that within spherical collapse
models (be it dust, tangential pressure
or others) the data set leading to naked singularities is not a
subset of `zero measure' of the set of all possible initial data.

Despite all this work we can still say that much more is to be
understood about the role that general pressures play during the final
stages of collapse. Perfect fluid collapse has been studied mostly
under some simplifying assumptions and restrictions in order to
gain an understanding, but a general formalism for perfect fluids
described by a physically valid equation of state is still lacking
due to the intrinsic difficulties arising from Einstein equations.
Considering both radial and tangential pressures is a fundamental
step in order to better understand what happens in the ultra-dense
regions that forms at the center of the collapsing cloud prior to
the formation of the singularity. For this reason, perfect fluids
appear as a natural choice since these are the models that are
commonly used to describe gravitating stars in equilibrium and since
it can be shown that near the center of the cloud regularity
implies that matter must behave like a perfect fluid.

In the present paper we use a general formalism to analyze
the structure of collapse in the presence of perfect fluid pressures.
This helps to understand better realistic collapse scenarios and their
outcomes and brings out clearly the role played by pressures towards
the formation of black holes or naked singularities as the endstate
of collapse. We examine what are the key features
that determine the final outcome of collapse in terms of a black hole
or a naked singularity when perfect fluids, without any restriction
imposed by the choice of an equation of state, are considered.
The reason we do not assume an explicit equation of state here
is that the behavior of matter in ultra-dense states in the final
stages of collapse is unknown.
On the other hand, having regularity and energy conditions
satisfied provides a physically reasonable framework to
study collapse endstates.

We find that not only naked singularities are not ruled out in
perfect fluid collapse scenarios but also that the separation between
the black hole region and the naked singularity region in the space
of all possible evolutions has some interesting features. In
particular, we show that the introduction of small pressures can
drastically change the final fate of the well-known pressureless
models. For example, we see that adding a small pressure
perturbation to an inhomogeneous dust model leading to a black hole
can be enough to change the outcome of collapse to a naked singularity,
and viceversa.

Further to this, we investigate here the space of initial
data and collapse evolutions in generality, in order to examine
the genericity of naked singularities in collapse. To study
small pressure perturbations as well as the genericity and
stability aspects, we use the general formalism for spherically
symmetric collapse developed earlier
\cite{ndim1}, \cite{ndim2}
in order to address the basic problem of how generic
is a given collapse scenario which leads to the formation of naked
singularities. Given the existence of an increasing number of
models describing collapse leading to a naked singularity, the
issue of genericity and stability of such models in the
space of initial data has become the crucial ingredient in order
to decide whether the Cosmic Censorship Hypothesis
in its present form should be conserved, modified or dropped
altogether.

It should be noted, however, that the concepts such as
genericity and stability are far from well-defined in a unique
manner in general relativity, as opposed to the Newtonian gravity.
A major difficulty towards such a task is the non-uniqueness
of topology, or the concept of `nearness' itself in a given
spacetime geometry
\cite{HE}.
One could define topology on a space of spacetime metrics by
requiring that the metric component values are `nearby' or also
additionally requiring that their $n$-th derivatives are also
nearby, and in each case the resulting topologies will be different.
This is in fact connected in a way with the basic problem in
arriving at a well-formulated statement of the cosmic censorship
itself. There have been attempts in the past to examine the genericity
and stability of naked singularities in special cases.
For example, in
\cite{C}
it was shown that for certain classes of
massless scalar field
collapse the initial data leading to naked singularity has, in
a certain sense, a positive codimension, and so the occurrence
of naked singularity is unstable in that sense. On the other
hand, it was shown in
\cite{Sar-Pramana}, \cite{SG}
that naked singularity occurrence is stable
in the sense of the data sets leading to the same being
open in the space of initial data.
But these need not be dense in this space and so `non-generic'
if we use the definition of `genericity' in the sense given
in the dynamical systems theory (where a set of initial data
leading to a certain outcome is said to be generic if it is
open and dense within the set of all initial data).
In that case, however, both black hole and naked singularity
final states turn out to be `non-generic'.

Therefore, in the following we adopt a more physical
definition of `genericity' in the sense of `abundance', and we
call generic an initial data set that has a
non-zero measure, and which is open (though not necessarily dense)
in the set of all initial data.
With this definition, the results
obtained in
\cite{Sar-Pramana} and \cite{SG}
would mean that both black hole and naked singularity
are generic collapse endstates. We note that
we do not deal here with the self-similar models, or scalar fields,
which is a somewhat special case. Therefore, the issue of
genericity and stability of naked singularities in collapse
remains wide open, for spherically symmetric as well as non-spherical
models and for different forms of matter fields. Our consideration
here treats in this respect a wide variety of physically
reasonable matter fields for spherically symmetric
gravitational collapse.

In section \ref{einstein}, the general structure for Einstein
equations to study spherical collapse is reviewed and we describe
how the equations can be integrated thus obtaining the equation of
motion for the system.
In Section \ref{perturbation} we examine the structure of the
initial data sets of collapse leading to black hole and naked
singularity to gain an insight on genericity of such outcomes for
some special models and effect of introducing small pressure
perturbations is investigated.
Section \ref{genericity} then considers the genericity aspects of
the outcomes of collapse with respect to initial data sets. We prove
that the initial data sets leading to black holes and naked
singularities in the space of all initial data sets for perfect
fluid collapse are both generic.
Section \ref{eos} is devoted to a brief discussion on equations
of state.
Finally, in section \ref{remarks}
we outline the key features of the above approach and its
advantages, and point to possible future uses of the same for
astrophysical and numerical applications.

\section{Dynamical evolution of collapse}\label{einstein}

In this section we summarize and review the key features
on spherical gravitational collapse analysis, and also reformulate
some of the key quantities and equations, especially those
relating to the nature and behaviour of the final singularity
curve. This will be useful in a later section in analyzing
the small pressure perturbations in a given collapse scenario,
and subsequently towards a general analysis of the genericity
aspects of the occurrence of naked singularities and black holes
as collapse final states.

The regularity conditions and energy conditions
that give physically reasonable models are discussed here.
The final stages of collapse are discussed, evaluating key elements
that determine when the outcome will be a black hole or a naked
singularity. We shall find a function, related to the tangent
of outgoing geodesics at the singularity whose sign solely
determines the time of formation of trapped surfaces in relation
with the time of formation of the singularity. We also analyze
here the occurrence of trapped surfaces during collapse and
the possibility that radial null geodesics do escape thus making
it visible. We see how both features are related to the sign
of the above mentioned function, thus obtaining a necessary
and sufficient condition for the visibility of the
singularity.

\subsection{Einstein equations}

The most general spacetime describing a spherically
symmetric collapsing cloud in comoving coordinates $r$ and $t$
depends upon three functions $\nu(r,t)$, $\psi(r,t)$ and $R(r,t)$,
and takes the form,
\begin{equation}\label{metric}
ds^2=-e^{2\nu(t, r)}dt^2+e^{2\psi(t, r)}dr^2+R(t, r)^2d\Omega^2 \; .
\end{equation}
The energy momentum tensor reads,
\begin{equation}
T_t^t=-\rho; \; T_r^r=p_r; \; T_\theta^\theta=T_\phi^\phi=p_\theta \; ,
\end{equation}
where $\rho$ is the energy density and $p_r$ and $p_\theta$
are the radial and tangential stresses. The metric functions
$\nu$, $\psi$ and $R$ are related to the energy-momentum tensor
via the Einstein equations that can be written in the form:
\begin{eqnarray}\label{p}
p_r&=&-\frac{\dot{F}}{R^2\dot{R}} \; ,\\ \label{rho}
\rho&=&\frac{F'}{R^2R'} \; ,\\ \label{nu}
\nu'&=&2\frac{p_\theta-p_r}{\rho+p_r}\frac{R'}{R}-\frac{p_r'}
{\rho+p_r} \; ,\\ \label{G}
2\dot{R}'&=&R'\frac{\dot{G}}{G}+\dot{R}\frac{H'}{H} \; ,\\
\label{F}
F&=&R(1-G+H) \; ,
\end{eqnarray}
where $F$ is the Misner-Sharp mass of the system
(representing the amount of matter enclosed in the comoving
shell labeled by $r$ at the time $t$) and for convenience we have
defined the functions $H$ and $G$ as,
\begin{equation} \label{HG}
H =e^{-2\nu(r, v)}\dot{R}^2 , \; G=e^{-2\psi(r, v)}R'^2.
\end{equation}
The collapse scenario is obtained by requiring $\dot{R}<0$
and the central `shell-focusing' singularity is achieved for
$R=0$, where the density and spacetime curvatures blow up.
Divergence of $\rho$ is obtained also whenever $R'=0$, thus
indicating the presence of a `shell-crossing' singularity. Such
singularities are generally believed to be gravitationally weak
and do not correspond to divergence of curvature scalars, therefore
indicating that they are removable by a suitable change of
coordinates
\cite{cross1},\cite{cross2}.
For this reason in the forthcoming discussion we will be
concerned only with the shell-focusing singularity
thus assuming $R'> 0$.

Since there is a scale invariance degree of freedom
we can choose the initial time in such a way so that $R(r, t_i)=r$.
Therefore we can introduce the scaling function $v(r,t)$
defined by,
\begin{equation}
    R=rv \; ,
\end{equation}
with $v(r,t_i)=1$, so that the collapse will be described by
$\dot{v}<0$ and the singularity will be reached at $v=0$. We see
that this is a better definition for the singularity since at
$v\neq 0$ the energy density does not diverge anywhere on the
spacelike surfaces, including at the center $r=0$. This is seen
immediately from the regular behaviour of the mass function near
the center that imposes that $F$ must go like $r^3$ close to $r=0$
(as it will be shown later). Such a regularity
was not clear from equation \eqref{rho}, especially along the
central curve $r=0$, where we have $R=0$, without the introduction
of $v$. In this manner the divergence of $\rho$ is only reached
at the singularity. We note that $v$ acts like a label for
successive events near the singularity since it is monotonically
decreasing in $t$ and therefore can be used as a `time' coordinate
in the place of $t$ itself near the singularity.

We shall consider the Misner-Sharp mass to be in general
a function of the comoving radius $r$ and the comoving time $t$,
expressed via the `temporal' label $v$ as
\begin{equation}
    F=F(r,v(r,t)) \; .
\end{equation}
Near the center of the cloud this is just equivalent
to a rewriting $F(r,t)$.

It can be shown that vanishing of the pressure gradients near
the center of the cloud imply that the radial and tangential stresses
must assume the same value in the limit of approach to the center
\cite{ndim1}.
This requirement comes from the fact that the metric functions
should be at least $\mathcal{C}^2$ at the center of the cloud and
is a consequence of the fact that the Einstein equation \eqref{nu}
for a general fluid contains a term in $p_r-p_\theta$ that
therefore must vanish at $r=0$.
Since we are interested in the final stages of collapse of the
core of a star it is therefore straightforward to take that
the cloud behaves like a perfect fluid in proximity of $r=0$.
We shall then take,
$$
p_r=p_\theta=p \; .
$$

In such a case, we are then left with five equations in
the six unknowns $\rho$, $p$, $F$, $\nu$, $\psi$ and $v$. The system
becomes closed once an equation of state for the fluid matter is
defined or assumed, but in general it is possible to study physically
valid dynamics (namely those satisfying regularity and energy
conditions) without assuming a priori an equation of state. In fact
it is reasonable to suppose that any equation of state that holds at
the departure from equilibrium, when the gravitational collapse
commences, will not continue to hold in the extreme regimes achieved
when approaching the singularity. In this case, we are then left
with the freedom to choose one of the functions.
If we take $F$ as the free function then from Einstein
equations \eqref{p} and \eqref{rho}, $p$ and $\rho$ will follow
immediately and they can be evaluated explicitly once we know
$v$ and its derivatives.

Further, from the requirement for perfect fluid collapse we
can write equation \eqref{nu} as
\begin{equation}
    \nu'=-\frac{p'}{\rho+p} \; .
\end{equation}
Equation \eqref{G} can be integrated once we define a
suitable function $A(r,v)$ from
\begin{equation}\label{defA}
    A_{,v}=\nu'\frac{r}{R'} \; .
\end{equation}
Then we get,
\begin{equation}
    G(r,v)=b(r)e^{2A(r,v)} \; ,
\end{equation}
where the integration function $b(r)$ can be interpreted
in analogy with the dust models and is seen to be related to
the velocity of the infalling matter shells.

Finally, from equation \eqref{F} we can write the equation
of motion for $R$ in the form of an effective potential as
\begin{equation}\label{Rdot}
    \dot{R}^2=e^{2\nu}\left(\frac{F}{R}+G-1\right) \; .
\end{equation}
This allows us to study the dynamics of the collapsing system
in analogy of the usual phase-space tools of the classical
mechanics type models.

Equation \eqref{Rdot} can be expressed in terms of the
scaling factor $v$ as,
\begin{equation}\label{vdot}
    \dot{v}=-e^\nu\sqrt{\frac{F}{r^3v}+\frac{be^{2A}-1}{r^2}} \; ,
\end{equation}
where the minus sign has been considered in order to
study the collapse. We see that in order to have a solution
we must have $\left(\frac{F}{r^3v}+\frac{be^{2A}-1}{r^2}\right)>0$,
we can call this a `reality condition' that is necessary for
the collapse dynamics to occur. Solving the equation \eqref{vdot}
solves the set of Einstein equations.

\subsection{Regularity and energy conditions}\label{conditions}

Einstein equations provide the relations between the
spacetime geometry and the matter distribution within it,
however, they do not give any statement about the type of
matter that is responsible for the geometry. On a physical ground,
not every type of matter distribution is allowed, and therefore
some restrictions on the possible matter models must be made
based on considerations of physical reasonableness. This
usually comes in the form of energy conditions ensuring the
positivity of mass-energy density.

Further, regularity conditions must be imposed
in order for the matter fields to be well-behaved at the
initial epoch from which the collapse evolves and at
the center of the cloud. Firstly, the
finiteness of the energy density at all times anteceding
the singularity and regularity of the Misner-Sharp
mass in $r=0$ imply that in general we must have,
\begin{equation}\label{mass}
    F(r,t)=r^3M(r,v(r,t)) \; ,
\end{equation}
where $M$ is a regular function going to a finite
value $M_0$ in the limit of approach to the center.
If $F$ does not go as $r^3$ or higher power in the limit of
approach to the center $r=0$, we immediately see from
the Einstein equation for $\rho$ that there would
be a singularity at the center at the initial epoch,
which is not allowed by the regularity conditions
as we are interested in collapse from regular initial
configurations. Also, requiring that the energy density has
no cusps at the center is reflected in the condition,
\begin{equation}
    M'(0,v)=0 \; .
\end{equation}

As seen before, the behaviour of the pressure
gradients near $r=0$ suggests that the tangential
and radial pressures become equal in limit of approach
to the center, thus justifying our assumption of
a perfect fluid type of matter.
Further since the gradient of the pressures must
vanish at $r=0$, we see that $p'\simeq r$ near $r=0$
which for the metric function $\nu$ implies that
near the center,
\begin{equation}
\nu(r,t)=r^2g(r,v(r,t))+g_0(t) \; ,
\end{equation}
where the function $g_0(t)$ can be absorbed in
a redefinition of the time coordinate $t$. From the above,
via equation \eqref{defA} we can write
\begin{equation}\label{av}
  A_{,v}=\frac{2g+rg'}{R'}r^2 \; .
\end{equation}

From the analogy with the Lemaitre-Tolman-Bondi (LTB) models
we can evaluate the regularity requirements for the
velocity profile $b(r)$. Since near the center $b$ can be
written as,
\begin{equation}
    b(r)=1+r^2b_0(r) \; ,
\end{equation}
we now see how to interpret the free function
$b(r)$ in relation with the known LTB
dust models. In fact the cases with $b_0$ constant
are equivalent to the bound ($b_0<0$), unbound ($b_0>0$)
and marginally bound ($b_0=0$) LTB collapse models.
Also, the condition that there be no shell-crossing
singularities may imply some further restrictions
on $b$.

As is known, matter models describing physically
realistic sources must be constrained by some energy conditions.
The weak energy condition in our case implies
\begin{equation}
    \rho\geq 0 , \; \rho+p\geq 0 \; .
\end{equation}
The first one is achieved whenever $F'>0$. In fact
from equation \eqref{rho} we see that positivity of $\rho$
is compatible with positivity of $R'$ only if $F'>0$.
Therefore we must have
\begin{equation}\label{wec1}
    3M+rM'\geq 0 \; ,
\end{equation}
that close to the center will be satisfied whenever
$M(0,v)\geq 0$.
Now from $M'=M_{,r}+M_{,v}v'$ we can rewrite $\rho$ as
\begin{equation}
    \rho=\frac{3M+rM_{,r}-M_{,v}v}{v^2R'}-p \; ,
\end{equation}
from which we see that the second weak energy condition
is satisfied whenever
\begin{equation}\label{wec2}
    3M+rM_{,r}-M_{,v}v\geq 0 \; .
\end{equation}
Finally the choice of a mass profile satisfying the
above equation allows us to rewrite the condition \eqref{wec1} as
\begin{equation}
    3M+rM_{,r}-M_{,v}v\geq -M_{,v}R' \; ,
\end{equation}
which is obviously satisfied if the pressure is positive
(since it implies $M_{,v}<0$) but can be satisfied also by
some choice of negative pressure profiles.

The dynamical evolution of collapse is entirely determined
once the initial data set is given
\cite{Initial1}-\cite{Initial4}.
Specifying the initial conditions consists in prescribing
the values of the three metric functions and of the
density and pressure profiles as functions of $r$ on the
initial time surface given by $t=t_i$.
This reduces to defining the following functions:
\begin{equation}\nonumber
    \rho(r, t_i)=\rho_i(r), \; p(r, t_i)=p_{i}(r),  \;
    R(r, t_i)=R_i(r), \; \nu(r, v(r, t_i))=\nu_i(r), \; \psi(r, t_i)=\psi_i(r).
\end{equation}
At the initial time the choice of the scale function $v$ is
such that $R_i=r$, furthermore, from $R'_i=1$ we get $v'(r, t_i)=0$.

Since the initial data must obey Einstein equations it
follows that not all of the initial value functions can be
chosen arbitrarily.  In fact the choice of the mass profile
together with Einstein equations is enough to specify the
four remaining functions.
From equations \eqref{rho} and \eqref{mass}, writing
\begin{equation}
    M_i(r)=M(r, v(r, t_i))=M(r, 1) \; ,
\end{equation}
we get
\begin{equation}
    \rho_i=3M_i(r)+rM_i'(r) \; ,
\end{equation}
while from equation \eqref{p} we get,
\begin{equation}
    p_i=-(M_{,v})_i \; .
\end{equation}
From equation \eqref{nu} we can write,
\begin{equation}
    \nu_i(r)=r^2g(r, v(r, t_i))=r^2g_i(r) \; .
\end{equation}
with $g_i(r)$ related to $M_i$ by
\begin{equation}
    2rg_i+r^2g'_i=-\frac{p_i'}{\rho_i+p_i} \; .
\end{equation}
In turn, $\nu_i$ can be related to the function $A$, defined by equation
\eqref{defA}, at the initial time via equation \eqref{av},
\begin{equation}
    A_{,v}(r, v(r, t_i))=(A_{,v})_i=2g_ir^2+g_i'r^3 \; .
\end{equation}
Finally, the initial condition for $\psi$ can be related
to the initial value of the function $A(r, t)$ from equation
\eqref{G} and equation \eqref{defA},
\begin{equation}
    A(r, v(r, t_i))=A_i(r)=-\psi_i-\frac{1}{2}\ln b(r) \; .
\end{equation}

Since we are studying the final stages of collapse and
the formation of black holes and naked singularities, we
must require the initial configuration to be not trapped.
This will allow for the formation of trapped surfaces during
collapse and therefore we must require
\begin{equation}
    \frac{F_i(r)}{R_i}=r^2M_i(r)< 1 \; ,
\end{equation}
from which we see how the choice of the initial matter
configuration $M_i$ is related to the initial boundary of
the collapsing cloud. Some restrictions on the choices of the
radial boundary must be made in order not to have trapped
surfaces at the initial time.
This condition is reflected on the initial configuration
for $G$ and $H$ since $1-\frac{F}{R}=G-H$ and this condition
also gives some constraints on the initial velocity.
In fact to avoid trapped surfaces at the
initial time the velocity of the infalling shells must satisfy
\begin{equation}
    |\dot{R}|>\sqrt{b}e^{A+\nu} \; .
\end{equation}
We see that the initial velocity of the cloud must
always be positive and that the case of equilibrium configuration
where $\dot{R}=0$ can be taken only in the limit.

The consideration of a
perfect fluid matter model implies that the Misner-Sharp mass is in
general not conserved during collapse. Therefore the matching
with an exterior spherically symmetric solution leads to
consider the generalized Vaidya spacetime.
It can be proven that matching to a generalized Vaidya
exterior is always possible when the collapsing cloud is taken
to have compact support within the boundary taken at $r=r_b$,
and the pressure of the matter is assumed to vanish at the
boundary
\cite{matching1}-\cite{matching4}.
Matching conditions imply continuity of the
metric and its first derivatives across the boundary surface.
Such a matching is in principle always possible but it
should be noted that matching conditions together with
regularity and energy conditions, might impose some further
restrictions on the allowed initial configurations.

In the following we are interested in the local
visibility of the central singularity occurring at the
end of the collapse. Therefore we will restrict our attention
to a neighborhood of the central line $r=0$.
In this case it is easy to see that there always exist
a neighborhood for which no shell crossing singularities occur.
This is seen by the fact that $R'=v+rv'$ and therefore, since
$v>0$ and shell crossing singularities are defined by $R'=0$,
we can always fulfill $R'>0$ in the vicinity of the center.
Furthermore, as it was mentioned before, matter behaves
like a perfect fluid close to the center. This can be seen
from the fact that regularity of the metric functions at
the initial time requires that $\nu'$ does not blow up at the
regular center. This in turn implies that from equation
\eqref{nu} we must have $p_r(0, t_i)-p_\theta(0, t_i)=0$,
and the condition holds for any time $t$ before the
singularity.

\subsection{Collapse final states }\label{collapse}

We study now the possible outcomes of collapse
evolution. It is known that in general the final fate of
the complete collapse of the matter cloud will be either
a black hole or a naked singularity, depending on the
choice of the initial data and the dynamical evolutions
as allowed by the Einstein equations.

In order to analyze the final outcome of collapse
we shall perform a change of coordinates from $(r,t)$ to $(r,v)$,
thus considering $t=t(r,v)$. As mentioned earlier this is
always possible near the center of the cloud due to the
monotonic behaviour of $v$. In this case the derivative of $v$
with respect to $r$ in the $(r,t)$ coordinates shall be considered
as a function of the new coordinates, $v'=w(r,v)$. We see that
regularity at the center of the cloud implies
$w\rightarrow 0$ as $r\rightarrow 0$.

We can then consider the metric function $\nu=\nu(r,v)$
which is given by equation \eqref{nu}, which now becomes,
\begin{equation}\label{nu'}
    \nu'=\frac{M_{,vr}v+(M_{,vv}v-2M_{,v})w}{(3M+rM_{,r}-M_{,v}v)v}R' \; ,
\end{equation}
where now $R'=v+rw$ and $_{,r}$ denotes derivative with
respect to $r$ in the $(r,v)$ coordinates. This implies,
\begin{equation}\label{A,v}
    A(r,v)=\int_v^1\frac{M_{,vr}v+(M_{,vv}v-2M_{,v})w}{(3M+rM_{,r}-M_{,v}v)v}rdv \; .
\end{equation}
For the sake of clarity, we may assume here that near
the center the mass function $M(r,v)$ can be written as a
series as,
\begin{equation}\label{powerM}
    M(r,v)=M_0(v)+M_1(v)r+M_2(v)r^2+ o(r^4) \; .
\end{equation}

As a regularity condition, we take $M_1=0$
and $A\simeq r^2$. The function $A$ can then be written as
an expansion and it takes the form,
\begin{equation}
    A(r,v)=A_2(v)r^2+A_3(v)r^3+o(r^4) \; ,
\end{equation}
with the first terms $A_i(v)$ given by
\begin{eqnarray}\label{A2}
  A_2 &=& \int_v^1\frac{2M_{2,v}+\left(M_{0,vv}-\frac{2M_{0,v}}{v}
  \right)w_{,r}}{3M_0-M_{0,v}v}dv \; , \\ \label{A3}
  A_3 &=& \frac{1}{2}\int_v^1 \frac{6M_{3,v}+\left(M_{0,vv}
  -\frac{2M_{0,v}}{v}\right)w_{,rr}}{3M_0-M_{0,v}v}dv \; .
\end{eqnarray}
If we restrict our analysis to constant $v$ surfaces then
in equation \eqref{A3} we can put $w$ and its derivatives to be
zero. On the other hand if we approach the singularity along
a generic curve we cannot neglect the terms in $w$.

The equation of motion \eqref{vdot} takes the form
\begin{equation}
    \dot{v}=-e^\nu\sqrt{\frac{M}{v}+\frac{be^{2A}-1}{r^2}} \; ,
\end{equation}
which can be inverted to give the function $t(r,v)$
that represents the time at which the comoving shell
labeled $r$ reaches the event $v$,
\begin{equation}
    t(r,v)=t_i+\int_v^1\frac{e^{-\nu}}{\sqrt{\frac{M}{v}
+\frac{be^{2A}-1}{r^2}}}dv \; .
\end{equation}
Then the time at which the shell labeled by $r$
becomes singular can be written as a singularity curve as
\begin{equation}\label{tsr}
    t_s(r)=t(r,0)=t_i+\int_0^1\frac{e^{-\nu}}{\sqrt{\frac{M}{v}
+\frac{be^{2A}-1}{r^2}}}dv \; .
\end{equation}
Regularity ensures that, in general, $t(r,v)$
is at least $\mathcal{C}^2$ near the singularity and
therefore can be expanded as,
\begin{equation}
    t(r,v)=t(0,v)+\chi_1(v)r+\chi_2(v)r^2+o(r^3) \; ,
\end{equation}
with
\begin{equation}\label{t0v}
    t(0,v)=t_i+\int_v^1\frac{1}{\sqrt{\frac{M_0}{v}+b_0(0)+2A_2}}dv \; ,
\end{equation}
and $\chi_1=\frac{dt}{dr}\rvert_{r=0}$ and
$\chi_2=\frac{1}{2}\frac{d^2t}{dr^2}\rvert_{r=0}$.

Of course the situations with discontinuities can
be analyzed as well with minor technical modifications
to the above formalism. In our case, assuming that $t(r,v)$ can
be expanded implies that the first two terms in the expansion
of $A(r,v)$ must vanish. As seen before vanishing of first
term is consistent with the regularity condition for
$\nu$ that follows from the pressure gradients at the
center, while vanishing of second term implies that $M_1$
must be a constant, which gives, in accordance with
the requirement that the energy density has no cusps
at the center, that $M_1=0$.
The singularity curve then takes the form
\begin{equation}\label{ts}
    t_s(r)=t_0+r\chi_1(0)+r^2\chi_2(0)+o(r^3) \; ,
\end{equation}
where $t_0=t(0,0)$ is the time at which the
central shell becomes singular.

By a simple calculation, retaining for the sake of
completeness all the terms in the expansions of $M$ and
expanding $\nu$ as $\nu=g_2(v)r^2+...$, we obtain
\begin{equation}\label{chi1}
    \chi_1(v)=-\frac{1}{2}\int_v^1 \frac{\frac{M_1}{v}+b_{01}
+2A_3}{\left(\frac{M_0}{v}+b_{00}+2A_2\right)^{\frac{3}{2}}}dv \; ,
\end{equation}
and
\begin{equation}\label{chi2}
    \chi_2(v)=\int_v^1\left[\frac{3}{8}\frac{(\frac{M_1}{v}
   +b_{01}+2A_3)^2}{\left(\frac{M_0}{v}+b_{00}+2A_2\right)^{\frac{5}{2}}}
   -\frac{g_2}{\sqrt{\frac{M_0}{v}+b_{00}+2A_2}}
   -\frac{1}{2}\frac{\frac{M_2}{v}+2A_4+2A_2^2
   +2b_{00}A_2+b_{02}}{\left(\frac{M_0}{v}+b_{00}+2A_2\right)^{\frac{3}{2}}}\right]dv \; ,
\end{equation}
where we have defined $b_{0j}=b_0^{(j)}(0)$.

\subsection{Trapped surfaces and outgoing null geodesics}\label{horizon}

The apparent horizon is the boundary of trapped surfaces
which in general is given by,
\begin{equation}
    g_{ij}R_{,i}R_{,j}=0 \; .
\end{equation}
In the case of spherical collapse the above equation
reduces to $G-H=0$, which together with equation \eqref{F}
gives,
\begin{equation}\label{ah}
    1-\frac{F}{R}=0 \; .
\end{equation}
This describes a curve $v_{ah}(r)$ given by
\begin{equation}
    v_{ah}=r^2M(r, v_{ah}) \; .
\end{equation}
Inversely, the apparent horizon curve can be
expressed as the curve $t_{ah}(r)$ which gives the time
at which the shell labeled by $r$ becomes trapped.

In the dust case, the condition $M=M(r)$ implies
that approaching the singularity the radius of the apparent
horizon must shrink to zero thus leaving $t_0$ as the
only point of the singularity curve that can in principle
be visible to far away observers. On the other hand, in the
perfect fluid case we note that models where the mass profile
has different dependence on $v$ will lead to totally different
structures for the apparent horizon and the trapped region.
Indeed in full generality there can be cases where non-central
singularities become visible. This is possible in the case
in which $M(r,v)$ goes to zero as $v$ goes to zero,
leaving $\frac{M}{v}$ bounded (see e.g.
\cite{patil}).

Presently, we are interested in the case where
only the central singularity would be visible.
In order to understand what are the features
relevant towards determining the visibility of the
singularity to external observers we can evaluate
the time curve of the apparent horizon in such cases as
\begin{equation}\label{t-ah}
    t_{ah}(r)=t_s(r)-\int_0^{v_{ah}(r)}\frac{e^{-\nu}}
{\sqrt{\frac{M}{v}+\frac{be^{2A}-1}{r^2}}}dv \; ,
\end{equation}
where $t_s(r)$ is the singularity time curve, whose
initial point is $t_0=t_s(0)$. Near $r=0$, equation
\eqref{t-ah} can be written in the form,
\begin{equation}
    t_{ah}(r)=t_0+\chi_1(v_{ah})r+\chi_2(v_{ah})r^2+o(r^3) \; ,
\end{equation}
from which we see how the presence of pressures affects
the time of the formation of the apparent horizon.
In fact, all the initial configurations that cause
$\chi_1$ (or $\chi_2$ in case that $\chi_1$ vanishes) to be
positive will cause the apparent horizon curve to be increasing,
and trapped surfaces to form at a later stage than the singularity,
thus leaving the possibility that null geodesics
escape from the central singularity.
By studying the equation for outgoing radial null geodesics
it is possible to determine that whenever the apparent horizon
is increasing in time at the singularity there will be families
of outgoing future directed null geodesics that reach outside
observers from the central singularity, at least locally.

It can be shown that positivity of the first non-null
coefficient $\chi_i(0)$ is a necessary and sufficient condition
for the visibility of the central singularity
\cite{ndim1}.
Nevertheless the scenario of collapse of a cloud composed
of perfect fluid offers some more intriguing possibilities.
In fact we can see from equation \eqref{ah} that whenever the mass
function $F$ goes to zero as collapse evolves it is possible
to delay the formation of trapped surfaces in such a way that
a portion of the singularity curve $t_s(r)$ becomes timelike.
This in turn leads to the possibility that non-central
shells are visible when they become singular
\cite{timelike},
thus introducing a new scenario that is not possible
for dust collapse. It is easy to verify that in order for
the mass function to be radiated away during the evolution
the pressure of the fluid must be negative at some point
before the formation of the singularity. Despite this seemingly
artificial feature negative pressures are worth investigating
as they could point to a breakdown of classical gravity
and could describe the occurrence of quantum effects
close to the formation of the singularity
\cite{evaporation}.

\section{Small pressure perturbations of collapse}\label{perturbation}

We will now use the general formalism developed above to
study how the outcomes of gravitational collapse, either in terms
of a black hole or naked singularity, are altered once an
arbitrarily small pressure perturbation in the initial data
set is introduced.

The Lemaitre-Tolman-Bondi (LTB) model
(\cite{LTB1}-\cite{LTB3})
for inhomogeneous dust and homogeneous perfect fluid is reviewed
describing necessary conditions for the LTB collapse scenario.
Then certain perfect fluid models are given, using the treatment above,
by making specific choices for the free function so that
it reduces to the LTB case for some values of the parameters.
We show how the choice of these parameters or introduction of small
pressure perturbations is enough to change the final outcome
of collapse of the inhomogeneous dust.

From equations \eqref{chi1} we see that if we account for regularity
and physically valid density and pressure profiles (typically
including only quadratic terms in $r$) we have $\chi_1=0$.
Then the final outcome of collapse will be decided by $\chi_2$ as
written in equation \eqref{chi2}. We see immediately that
once the matter model is fixed globally, thus specifying
$M$, the sign of $\chi_2$ depends continuously on the values
at the initial time taken by the parameters $M_2$, $b_{02}$
and $A_j$ (with $j=2,4$). By continuity then we can say that,
away from the critical surface for which $\chi_2=0$, if a certain
initial configuration leads to a black hole (thus having $\chi_2<0$),
then changing slightly the values of the initial
parameters $M$, $b_{0}$, $A$ will not change the final outcome.
The same result holds for naked singularities and leads us
to conclude that every initial data set for which $\chi_2 > 0$
will have a small neighborhood leading to the same outcome
\cite{Pramana}.
The same, however, cannot be said for the surface
separating these two possible outcomes of collapse, where
$\chi_2=0$. In this case it is the sign of the next
non-vanishing $\chi_i$ that determines the final outcome and
it is easy to see that the introduction of a small pressure such
that $\chi_j$ becomes non-zero for some $j<i$ can change the
final fate from black hole to naked singularity
and viceversa.

Consider the scenario where the coefficients
$\chi_i$ vanish for every $i$.
This critical surface represents the case of simultaneous
collapse, or when $t_s(r)=const$,
where a black hole forms at the end of collapse
and it includes (though it is not uniquely restricted to)
the Oppenheimer-Snyder-Datt homogeneous dust collapse model.
While this is the case for homogeneous dust, it is also
easy to show that for inhomogeneous dust and for perfect
fluid collapse also there are initial configurations that
lead to simultaneous collapse, once inhomogeneities,
velocity profile, and pressure are chosen suitably.

In fact if we consider collapse of general type I matter
fields, we can always suitably tune the parameters in order to
have simultaneous collapse and therefore a black hole final
outcome. Nevertheless, in all these cases, the introduction of
a small pressure can drastically change the final outcome by
making some $\chi_i$ turn positive.
Of course in full generality there will also be regions
in which the `reality condition' is not satisfied and therefore
no final outcome is possible. But if we restrict our attention
to a close neighborhood of the center we will always have a
complete collapse of the inner shells thus leading to a black hole
or a naked singularity.

In this sense we can consider a small perturbation of
any type I fluid collapse. We see that the initial data not
lying on the critical surface will not change the outcome of
collapse once a small perturbation in $M$ or $p$ or $b$ is
introduced. On the other hand, those initial data sets that belong
to the critical surface might indeed change outcome entirely
as a result of the introduction of a small inhomogeneity, or
a small pressure or small velocity.
We shall now consider below some collapse models
that can be obtained from the above formalism, and analyze
these under the introduction of small pressure
perturbations.

\subsection{Lemaitre-Tolman-Bondi collapse}

The simplest model that can be studied for small pressure
perturbations is the well-known Lemaitre-Tolman-Bondi spacetime,
where the matter form is dust with pressures assumed to be
vanishing. It is interesting to know how the collapse outcome
would change when small pressure perturbations are introduced
in the cloud, which is a more realistic scenario compared
to pressureless dust.
The spacetime metric in this case takes the form,
\begin{equation}
    ds^2=-dt^2+\frac{R'}{1+r^2b_0(r)}dr^2+R^2d\Omega^2 \; ,
\end{equation}
and it can be obtained from the above formalism not only if we
impose the matter to be dust but also once we require homogeneity
of the pressures (namely imposing $p'=0$).
In fact, if we take $p=p(t)$ or $p=0$, from Einstein equations,
together with the regularity condition for $\nu$ and $b(r)$
we obtain $\nu = 0$ and $G=1+r^2b_0(r)$. The equation of motion
\eqref{vdot} becomes $\dot{v} = \sqrt{\frac{M}{v}+b_0}$ where
$M$ is a function of $t$ only in the case of homogeneous
pressure, and it is a function of $r$ only in the case of dust.
In the case where $M=M_0$ is a constant we retrieve the
Oppenheimer-Snyder-Datt homogeneous dust model that, as it is
known, develops a black hole at the end of collapse.

\begin{figure}[hh]
\includegraphics[scale=1]{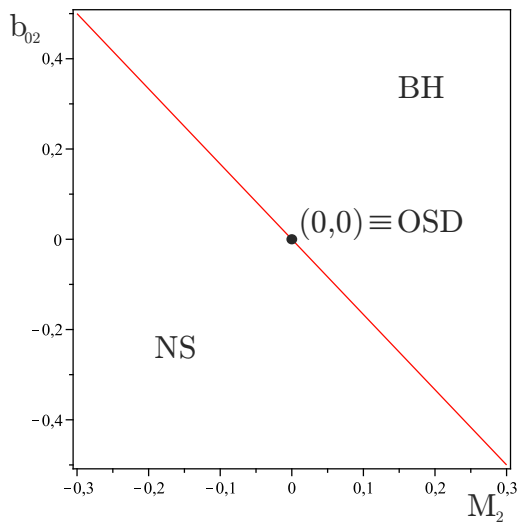}
\caption{The phase space of initial data for the LTB collapse model
with $M_0=1$ and $M_1=b_{00}=b_{01}=0$. In this case $\chi_1=0$ and $\chi_2$
determines the final outcome of collapse depending on the values of $M_2$ and $b_{02}$.
There are initial data sets that have a whole neighborhood
leading to the same outcome in terms of either a black hole
or naked singularity. The OSD case lies on
the critical surface separating the two outcomes which is defined by
$\chi_i=0$ for all $i$.}
\label{fig1}
\end{figure}

The inhomogeneous dust model is obtained by requiring
$M=M(r)$. In this case from equation \eqref{p} follows $p=0$
and in general $v$ can be a function of $r$ and $t$ (requiring
$v=v(t)$ is a necessary and sufficient condition to obtain
the OSD case). The final outcome of collapse
is fully determined once the mass profile and the velocity
profile are assigned (see figure \ref{fig1}).

For example, in the marginally bound case (namely
$b_0=0$) the singularity curve for inhomogeneous dust
becomes $t_s(r)=t_i+\frac{2}{3}\frac{1}{\sqrt{M(r)}}$ and the
apparent horizon curve is given by $t_{ah}(r)=t_s(r)-\frac{2}{3}r^2M(r)$
and in general collapse may lead to black hole or naked
singularity depending on the behaviour of the mass profile
$M(r)$. By writing $M(r)$ near $r=0$ as a series
$M(r)=M_0+M_1r+M_2r^2+o(r^3)$ we see that the lowest order
non-vanishing $M_i$ (with $i>0$) governs the final outcome
of collapse. As expected in this case, we have
\begin{eqnarray}
  \chi_1(0) &=& -\frac{1}{3}\frac{M_1}{M_0^{\frac{3}{2}}} \; ,\\
  \chi_2(0) &=& \frac{1}{4}\frac{M_1^2}
{M_0^{\frac{5}{2}}}-\frac{1}{3}\frac{M_2}{M_0^{\frac{3}{2}}} \; .
\end{eqnarray}

In the perfect fluid LTB model (corresponding to the FRW
cosmological models in case of expansion) we have that requiring
$v=v(t)$ is a sufficient condition for having homogeneous
collapse. In fact the following two statement can be easily proved:
\begin{enumerate}
  \item $v=v(t) \Leftrightarrow \begin{cases}
M(r,t)=M(t) \; , \\ b_0(r)=k \; .\end{cases}$
  \item $v=v(t) \Rightarrow \begin{cases} \rho=\rho(t)
\; ,\\ p=p(t) \; .\end{cases}$
\end{enumerate}

The overall behaviour of the collapsing cloud is
determined by the three functions $M(r,t)$, $v(r,t)$ and
$b_0(r)$ (which, as we have seen, are not independent from one
another) and the special cases of Oppenheimer-Snyder-Datt metric
and Lemaitre-Tolman-Bondi perfect fluid metric can be
summarized as follows.
\begin{equation}\nonumber
  \begin{array}{ccccc}
  M_0, & k, & v(t) & \Leftrightarrow & \text{Homogeneous dust} \\
  M(r), & b_0(r), & v(r,t) & \Leftrightarrow & \text{Inhomogeneous dust} \\
  M(t), & k, & v(t) & \Rightarrow & \text{Homogeneous perfect fluid}
  \end{array}
\end{equation}

\subsection{Perturbation of inhomogeneous dust}

We consider now an example based on the above framework
by introducing a small pressure perturbation to the inhomogeneous
dust model described in the previous section. We consider in
general $v=v(r,t)$ and the mass function is chosen
of the form
\begin{equation}
    M=M_0+M_2(v)r^2 \; ,
\end{equation}
where $M_0$ is a constant and the pressure perturbation
is taken to be small in the sense that $M_0 \gg |M_2|$
at all times (in this way, as collapse progresses and the
density diverges the model remains close to the LTB collapse as
the pressure is always smaller than the density).
We immediately see that setting $M_2=C$
reduces the model to that of inhomogeneous dust (and $C=0$ further
gives the OSD homogeneous dust).
We note that in this case no non-central singularities
are visible since $M$ does not vanish at $v=0$. Therefore, just
like in the dust case, only the central singularity at $r=0$
might eventually be visible.

In this case we can integrate equation \eqref{A2}
explicitly to obtain
\begin{equation}
    A_2(v)=\frac{2}{3}\frac{M_2(v)-C}{M_0} \, ,
\end{equation}
where $C$ now is the value taken by $M_2$ at the initial
time when $v=1$. Therefore we can take the mass function in
such a way that it corresponds to the LTB inhomogeneous dust
at initial time with the pressure perturbation being triggered
only at a later stage. We can therefore take,
\begin{equation}
    M_2(v)=C+\epsilon(v) \; ,
\end{equation}
and the initial condition $M_2(1)=C$ implies that
$\epsilon(1)=0$ (remember that $v\in[0,1]$).
By taking all the higher order terms to be vanishing we
easily see that $A_i=0$ for $i>2$ and
\begin{equation}
    g_0=\frac{1}{2}A_{2,v}v=\frac{1}{3}\frac{M_{2,v}v}{M_0} \; ,
\end{equation}
near $r=0$.

In this case the pressure and the energy density near $r=0$ become
\begin{equation}
    p=-\frac{\epsilon_{,v}}{v^2}r^2, \; \rho=\frac{3M_0}{v^3}+5\frac{C+\epsilon}{v^3}r^2 \; .
\end{equation}
We therefore have two simple possibilities for the choice
of the free function $\epsilon$ which determines $M$:
\begin{enumerate}
  \item $\epsilon>0$ which implies $\epsilon_{,v}<0$ and positive pressure.
  \item $\epsilon<0$ which implies $\epsilon_{,v}>0$ and negative pressure.
\end{enumerate}

From the above, further assuming $b_0=0$ for simplicity
in accordance with marginally bound LTB models, it follows
immediately that $\chi_1(0)=0$ and
\begin{equation}
    \chi_2(0)=-\frac{1}{2}\int^1_0\frac{(C+H(v))\sqrt{v}}{\left(M_0+\frac{4}{3}\frac{\epsilon v}{M_0}\right)^{\frac{3}{2}}}dv
    -\frac{4}{9M_0^2}\int^1_0 \frac{\epsilon v(\epsilon+\epsilon_{,v}v)\sqrt{v}}{\left(M_0+\frac{4}{3}\frac{\epsilon v}{M_0}\right)^{\frac{3}{2}}}dv \; ,
\end{equation}
where we defined the function
\begin{equation}
    H(v)=\left(\epsilon+\frac{2}{3}\epsilon_{,v}v\right) \; .
\end{equation}
We see that $\chi_2$ is divided into two integrals. If
we assume that the pressure perturbation is small (thus considering
$M_0$ to be big) then the second integral can in principle be
neglected. In fact for a suitable choice of $M_0$ it is not difficult
to prove that the function at the denominator will be positive
and monotonically increasing, and therefore it would not
affect the sign of the integral, while the second integral will
be small enough as compared to the first one. Positivity of
$\chi_2$ will then be decided by the sign of $C+H(v)$.

\begin{figure}[hh]
\begin{minipage}[b]{0.4\linewidth}
\includegraphics[width=\textwidth]{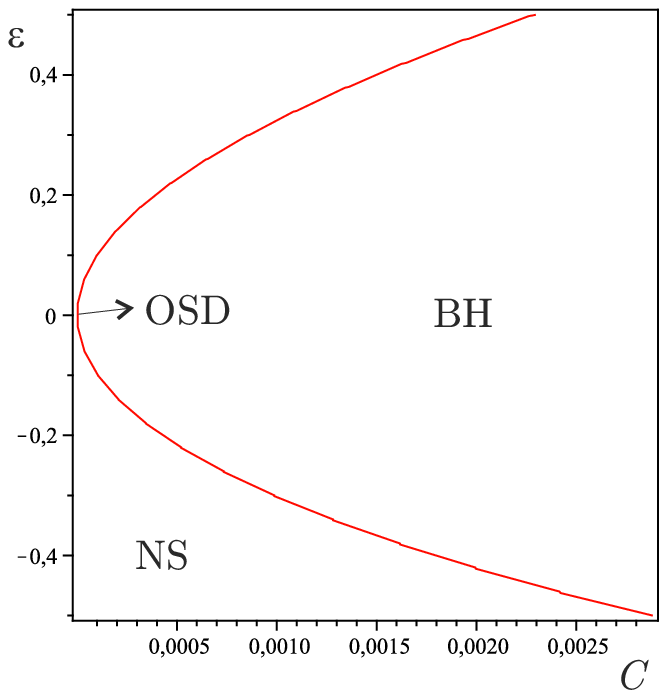}
\end{minipage}
\hspace{0.5cm}
\begin{minipage}[b]{0.4\linewidth}
\includegraphics[width=\textwidth]{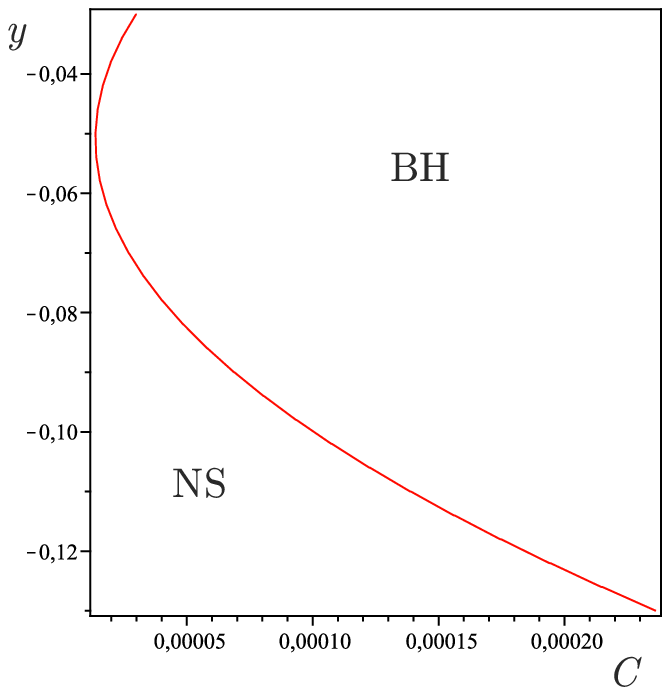}
\end{minipage}
\caption{The phase space of initial
data for perfect fluid collapse with $M_0=1$, $b_0=0$ and
$M=M_0+(C+\epsilon(v))r^2$. On the left the pressure is taken to be
$p=\varepsilon(3-\frac{2}{v}-\frac{1}{v^2})r^2$ and it reduces to zero for $\varepsilon=0$.
On the right the pressure perturbation is
given by $p(r,v)=\left(0.3+\frac{0.2}{v}y-\frac{0.1}{v}\right)r^2$.
The introduction of the pressure can uncover an otherwise clothed
singularity depending on the values of $\varepsilon$, $y$ and $C$.
Different choices of $M$ with $\epsilon(v)=y_0+y_1v+y_2v^2+y_3v^3$
will have a similar qualitative behaviour.}
\label{fig2}
\end{figure}

In order to have naked singularity we must have $\chi_2>0$.
This is certainly the case whenever
\begin{equation}
    C<-H(v) \; ,
\end{equation}
for any $v$. Therefore if we define $D_1=\min\{H(v), v \in[0,1]\}$
all the values of $C<D_1$ will lead to a naked singularity.
On the other hand, values of $C>D_2=\max\{H(v), v \in[0,1]\}$ will
lead to the formation of a black hole. For $C\in [D_1,D_2]$ the
explicit form of $\epsilon(v)$ is what determines the sign of
$\chi_2$. It can be proven that we can have models in which positive
values of $C$ lead to the formation of a naked singularity
whereas the inhomogeneous dust case led to a black hole.

\subsection{A perturbation with pressure $\bf{p(r)}$}

We now analyze another case where the pressure perturbation
introduced does not
depend explicitly on $v$. A similar situation was investigated
by one of us earlier in
\cite{Sar-Pramana}.
Here we consider a pressure perturbation of LTB models
of the form $p=p(r)$.

If we impose
\begin{equation}
    M_{,vv}v-2M_{,v}=0 \; ,
\end{equation}
in the equation \eqref{nu'}, we can then solve for $M$
and explicitly evaluate $G$.
We obtain in that case,
\begin{equation}
    M(r,v)=y(r)v^3+z(r) \; ,
\end{equation}
where $y(r)$ and $z(r)$ come from integration.
Here the case where $y=0$ reduces the system to dust.
In this sense, if we keep $|y|\ll z$ we will consider this model
to be a small perturbation of LTB in a similar way as
was discussed in the previous example. Then the radial
derivatives of $z$ will correspond to the inhomogeneities
in the LTB models while the pressure, or the function $y$,
can be taken to be either positive or negative (with positive
pressure corresponding to an increasing mass function
$M$ and negative pressure corresponding to a decreasing
mass function).

In order to work on a specific model for the sake
of clarity, we assume that $y$ and $z$ can be expanded near
the center as,
\begin{eqnarray}
  y(r) &=& y_0+y_1r+y_2r^2+...  \; ,\\
  z(r) &=& z_0+z_1r+z_2r^2+... \; .
\end{eqnarray}
Then the pressure and density become
\begin{equation}
  p = -3y(r), \;
  \rho = \frac{3z+z_{,r}r+y_{,r}rv^3}{v^2R'}-p \; .
\end{equation}
Imposing regularity requires that $y_{,r}(0)=z_{,r}(0)=0$,
which implies $y_1=z_1=0$. At the center of the cloud the
pressure and density become $p_0 = -3y_0$ and
$\rho_0 = 3\left(\frac{z_0}{v^3}+y_0\right)$, and the energy
conditions impose that $z_0\geq 0$ and $y_0\geq -z_0$.
Then from equation \eqref{A,v} we can integrate
explicitly to obtain,
\begin{equation}
    A(r,v)=\ln\left(\frac{u(r)+1}{u(r)+v^3}\right) \; ,
\end{equation}
where we have defined
\begin{equation}
    u(r)\equiv\frac{3z(r)+z_{,r}r}{y_{,r}r} \; .
\end{equation}
Then we get,
\begin{equation}
    G(r,v)=b(r)\left(\frac{u+1}{u+v^3}\right)^2 \; .
\end{equation}

\begin{figure}[hh]
\includegraphics[scale=1]{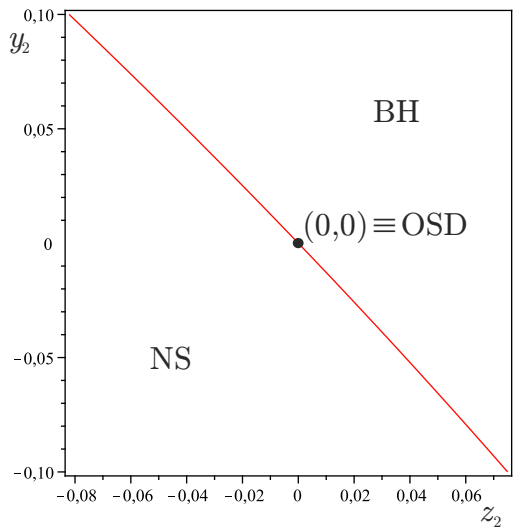}
\caption{The numerical evaluation of the sign of $\chi_2$ as a function of $y_2$ and $z_2$ provides
the phase space of initial data for the perfect
fluid model with $p=p(r)$. Here we have taken $z_0=1$ and $y_0=-1/2$.
Again, there are initial data sets
that have a whole neighborhood leading to the
same outcome which is either a black hole or naked singularity.
The OSD case lies on the critical
surface separating the black holes from naked singularities.}
\label{fig3}
\end{figure}

From the above expressions, we can easily obtain
now $\chi_1$ and $\chi_2$. For simplicity we consider here
the case where $b_0=y_4=0$. Then $\chi_1=0$ and
we get,
\begin{equation}
    \chi_2=-\frac{1}{2}\int^1_0\left[\frac{2\frac{y_2}{z_0}v^6}
    {\sqrt{y_0v^3+z_0+\frac{4}{3}\frac{y_2}{z_0}v(1-v^3)}}+
    \frac{y_2v^3+z_2
    +\frac{1}{3}\frac{y_2}{z_0^2}v(v^3-1)\left[y_2(v^3-\frac{1}{3})+\frac{5}{3}z_2\right]}
    {\left(y_0v^3+z_0+\frac{4}{3}\frac{y_2}{z_0}v(1-v^3)\right)
^{\frac{3}{2}}}\right]\sqrt{v}dv
\end{equation}
We see from here that the sign of $\chi_2$ is explicitly
determined
by the inhomogeneities ($z_2$) and the pressure
gradient ($y_2$) (see Fig.\ref{fig3}).

Once again taking $y_0=y_2=0$ and $z_2=0$ reduces the system
to the OSD collapse scenario and we see that the introduction
of the slightest pressure can change drastically the outcome
of collapse. On the other hand, taking only $y_0=y_2=0$ (with $z_2\neq 0$)
we retrieve the LTB model and once again to
change the final outcome of collapse we must choose $y_2$
suitably to balance the contribution to $\chi_2$
given by the inhomogeneities. Therefore we see again that
also within this perturbation model any collapse with initial data
taken in a neighborhood of a model leading to a certain outcome
and not lying on the critical surface will result in the same endstate.

The equation for the apparent horizon curve can be easily
written in this case and becomes
\begin{equation}
    r^2y(r)v_{ah}(r)^3-v_{ah}(r)+r^2z(r)=0 \; ,
\end{equation}
which is a cubic equation in $v_{ah}$ that admits in general
three solutions in the case where
$27r^4z^2-\frac{4}{r^2y}\geq 0$. Obviously this condition is
satisfied near $r=0$ for positive pressures (that correspond to
negative $y$) and this indicates, as already stated, that
the mass function is not vanishing at any time and therefore
the central shell becomes trapped at the time of formation
of the singularity. On the other hand, for negative pressures
the central shell is not trapped and the formation of the
apparent horizon can be shifted to some outer shells
or removed altogether.

\subsection{Perturbation of a general simultaneous collapse}

As we mentioned, the case of simultaneous collapse,
which means that the final state of collapse is necessarily
a black hole, need not be restricted to the Oppenheimer-Snyder-Datt
model only.
In fact for different kinds of general type I matter fields
there might be suitable choices of the parameters that lead
the final state of collapse to be simultaneous. This is of
course the case when the pressures are homogeneous, that is
represented by the time reversal of the Friedman-Robertson-Walker
model, but more general matter models might also lead to the
same behaviour.

Simultaneous collapse means that all matter shells terminate
into the singularity at the same time. Then we see from equation
\eqref{ts}, which describes the singularity curve, that all
coefficients $\chi_i(0)$ must vanish, or equivalently that
$t_s(r)=t_0$. From equation \eqref{tsr} we can see that
a sufficient condition for simultaneous collapse is
\begin{equation}\label{sim-b}
    \frac{e^{-2\nu}}{\frac{M}{v}+\frac{be^{2A}-1}{r^2}}=h(v) \; .
\end{equation}
This condition, for any given choice of the matter model,
leads to a choice of the free function $b(r)$ as
\begin{equation}
    b(r)= e^{-2A}\left(1+\frac{e^{-2\nu}}{h}r^2-\frac{M}{v}r^2\right) \; .
\end{equation}
It is easy to check that in the case of dust this reduces to
\begin{equation}
    b_0(r)=\frac{1}{h(v)}-\frac{M(r)}{v} \; ,
\end{equation}
and since the mass profile in this case is a function of
$r$ only, we conclude that equation \eqref{sim-b} is satisfied
for dust only by the case of homogeneous dust collapse
(where $M(r)=M_0$ and $b_0=k$). Nevertheless, as we have said,
this need not be the only case when the collapse is
simultaneous.
For example it's straightforward to see that when pressures
are considered, the same condition as
above holds for collapse of an homogeneous perfect fluid,
where, in this case, $M=M(t)$. Furthermore it is possible
that the condition \eqref{sim-b} can be satisfied by some suitable
function $b(r)$ also for more general pressure profiles,
since in general the mass function depends on both $r$ and $t$,
or for some other suitable choice of the velocity profile.

In order to better understand the conditions under which
we can have simultaneous collapse in full generality, let us now
consider a general perfect fluid matter model given by a choice
of $M(r,v)$ (which implies $A$ from equation \eqref{A,v}),
thus specifying all coefficients $M_i(v)$.
Firstly, we notice from equation \eqref{t0v} that
a suitable choice of $b_{00}$ is necessary in order for the
reality condition to be fulfilled near the center. Therefore,
once we made this choice and carried out the integration for
equation \eqref{chi1} we see that $\chi_1(0)$ depends linearly
on $b_{01}$ only. In fact we can write
\begin{equation}
    \chi_1(0)=-\frac{1}{2}\int_0^1\alpha_1(v)dv-
\frac{b_{01}}{2}\int^1_0\beta_1(v)dv \; ,
\end{equation}
from which we see that it will always be possible to choose
$b_{01}$ suitably such that $\chi_1(0)=0$. The same reasoning can
then be applied for all other coefficients $\chi_i(0)$ that will
depend linearly on $b_{0i}$ as,
\begin{equation}
    \chi_i(0)=\int_0^1\alpha_i(v)dv+b_{0i}\int^1_0\beta_i(v)dv \; ,
\end{equation}
and therefore for a given mass profile $M(r,v)$ we can have
simultaneous collapse if a suitable velocity profile $b(r)$ given by
\begin{equation}
    b(r)=1+r^2\sum_{i=0}^\infty\frac{b_{0i}}{i!}r^i \; ,
\end{equation}
exists. This means that the power series
$\sum_{i=0}^\infty\frac{b_{0i}}{i!}r^i$ should converge to some
function $b_0(r)$ with a radius of convergence greater than the
boundary of the cloud.

This is certainly possible in the case of homogeneous
pressures, where the condition that all $\chi_i(0)$ vanish imposes
that $\sum_{i=0}^\infty\frac{b_{0i}}{i!}r^i=\frac{C}{r2}$,
and therefore $b(r)=const$. Also, as we have seen in the examples
above, this might be possible for other matter profiles as well.
Given any such model leading to simultaneous collapse (and
thus to the formation of a black hole), we have shown that the
introduction of the slightest pressure perturbation in the initial
data can turn the final outcome into a naked singularity.

\begin{figure}[ht]
\begin{minipage}[b]{0.4\linewidth}
\includegraphics[width=\textwidth]{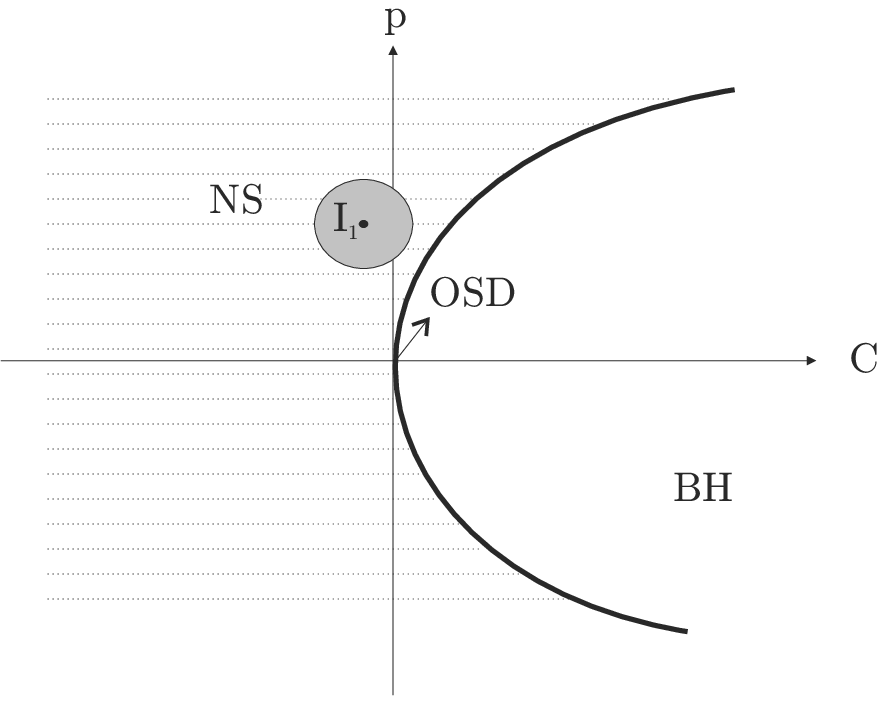}
\end{minipage}
\hspace{0.5cm}
\begin{minipage}[b]{0.4\linewidth}
\includegraphics[width=\textwidth]{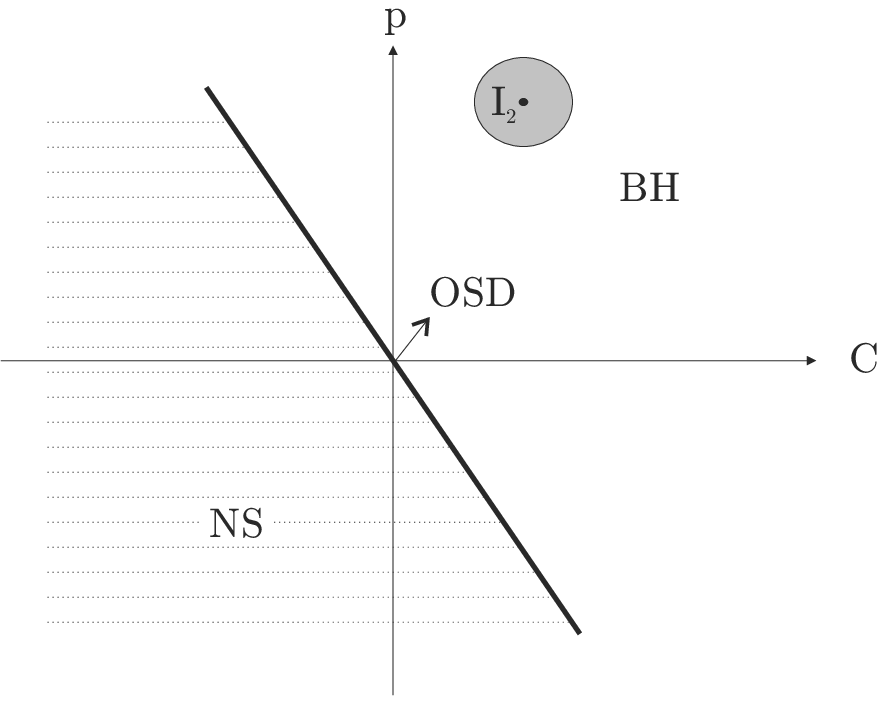}
\end{minipage}
\caption{Illustrative representation of the phase spaces of initial
data for different perfect fluid collapse models with assigned mass
profiles $M(r,v)$ and with $b_0=0$.
The introduction of pressures $p$ and inhomogeneities $C$
can uncover an otherwise clothed
singularity. Also, different choices of $M$ can
have the opposite effect, that is, a singularity
that was naked can become covered. The initial data set $\emph{I}_1$ has
a whole neighborhood lying in the initial data space leading to naked singularity
and can therefore be considered stable. The same holds for the initial data set
$\emph{I}_2$ leading to black hole. The LTB model (obtained for $p=0$ and)
can lead to either of the outcomes and the OSD scenario (obtained when there are
no inhomogeneities) always lies on the critical surface separating the two outcomes and
is therefore `unstable'.}
\label{fig4}
\end{figure}

Overall we have seen that the sign of $\chi_2$ and therefore the final
outcome of collapse shares similar qualitative behaviour
in different perfect fluid models as it is summarized in Fig. \ref{fig4}.

\section{Genericity of black holes and naked singularities} \label{genericity}
As we have seen, the final outcome of collapse depends
upon the evolution of the pressure, the density and the velocity
profiles. If the system is closed, as it is in the case where
an equation of state describing the relation between $p$ and
$\rho$ is given, then specifying the values of the above quantities
at the initial time uniquely determines the final outcome of
collapse. If the system is not closed, then we must further
specify the behaviour of the free functions. Once again for
every possible choice of the free function(s) the final outcome
of collapse is decided by the initial values of $p$,
$\rho$ and $b$.

The genericity is defined here as every
point in the initial data set leading to a naked singularity
(or a black hole) has a neighborhood in the space of initial data
for collapse whose points all lead to the same outcome. We show that
the initial data set leading the collapse to a naked singularity
forms an open subset of a suitable function space comprising of the
initial data, with respect to an appropriate norm which makes the
function space an infinite dimensional Banach space. The measure
theoretic aspects of this open set are considered and we argue that
a suitable well-defined measure of this set must be strictly
positive. This ensures genericity of initial data in a well-defined
manner.

At this point, the question of whether the given
outcome is `generic' or not in a certain suitable sense yet
to be defined, with respect to the allowed initial data sets,
arises naturally. We shall therefore analyze the expression
\eqref{chi1} for the genericity of initial data leading
the collapse to a naked singularity. Similar conclusions
apply to the case where $\chi_1$ vanishes and we must
analyze equation \eqref{chi2} and they
can be used to investigate the genericity of the
black hole formation scenario just as well.

As is known the concept of `genericity' is not well-defined
in General Relativity.
Normally, by the word `generic', one means `in abundance' or
`substantially big'.
This terminology has been used by many researcher working
in relativity, and in
gravitational collapse in particular (see for example,
\cite{genericity1}-\cite{genericity3}).
In the theory of Dynamical Systems, however, the definition
of `genericity' is given more tightly. There, one considers the
class $\mathrm{V}$ of all $\mathcal{C}^r$
vector fields (dynamical  systems) defined on a given manifold.
A property $\mathrm{P}$ satisfied by a vector field $X$ in
$\mathrm{V}$ is called
generic if the set of all vector fields satisfying this
property contains an open and dense subset of $\mathrm{V}$.
This was the definition used by one of us in
\cite{Sar-Pramana} and \cite{SG}. However, such a definition
would render both black holes and naked singularities to
be `non-generic', as we remarked earlier.
Therefore in the present paper we have opted for a less
stringent but physically more meaningful definition of genericity
by requiring that the subset is open, and that it has a
non-zero measure.
The main reason for this comes from the fact that the
`denseness' property depends on the topology used and the
parent space used, and there are no unique and unambiguous
definitions available in this regard as discussed earlier.
Hence the nomenclature of `generic' in the present paper
is used in the sense in which most of the relativists use it,
{\it i.e.} in the sense of abundance. This looks physically
more satisfactory definition, allowing both black holes
and naked singularities to be generic. In any case, the key
point is that regardless of the definition used, both
the collapse outcomes do share the same `genericity'
properties, which is what our work here shows.

\subsection{Existence of the set of initial data leading to naked singularity}

First of all we note that the functions must satisfy
the `reality condition' for the gravitational collapse to
take place, namely
\begin{equation}\label{GrindEQ}
    \frac{M_0(v)}{v} +b_{0} (0)+2a(0,v)>0 \; ,
\end{equation}
where we have defined $a(r,v)\equiv\frac{A(r,v)}{r^2}$.

We shall now prove that, given a mass function
$M(r,v)$, and the function $a(r,v)$ on the initial surface,
there exists a large class of velocity distribution functions
$b_0(r)$ such that the final outcome is a naked singularity.
We choose $b_0(r)$ to satisfy the following differential equation on
a constant $v$-surface,
\begin{equation} \label{GrindEQ1}
\frac{1}{2} \frac{\frac{M'(r,v)}{v} +b_{0} ^{{'} } (r)
+2a '(r,v)}{\left[\frac{M(r,v)}{v} +b_{0} (r)
+2a (r,v)\right]^{\frac{3}{2} } } = B(r,v) \; ,
\end{equation}
for $0\le r\le r_{b} $, where $B(r,v)$ is a continuous
function defined on a domain $D = [0 ,  r_b ] \times  [0 ,  1 ]$
such that
\begin{equation} \label{GrindEQ2}
B(0,v)=\frac{1}{2} \frac{\frac{M'(0,v)}{v} + b_{0} ^{{'} }
(0)+2a'(0,v)}{\left[\frac{M(0,v)}{v} +b_{0} (0)+2a
(0,v)\right]^{\frac{3}{2} } } <0 \; ,
\end{equation}
for all $v \in [0, 1 ]$. It will then follow that
\begin{equation} \label{GrindEQ3}
\chi_1 (0)=\lim_{v\to 0} \chi_1 (v)=-\int _{0}^{1}B(0,v)dv >0 \; .
\end{equation}
As seen above, this condition ensures that central
shell-focusing singularity will be naked.

We prove the existence of $b_0 (r)$ as a solution
of the differential equation \eqref{GrindEQ1} with initial
condition \eqref{GrindEQ2} which $B(r,v)$ will satisfy.
For this purpose, we define
\begin{equation} \label{GrindEQ4}
 x(r,v)=M(r,v)+vb_{0} (r)+2va(r,v) \; .
\end{equation}
Then equation \eqref{GrindEQ1} can be written as
\begin{equation}
  \frac{\sqrt{v}}{2}  \frac{  \left[M'(r,v)+vb_{0}' (r)
+2va'(r,v)\right]}{ \left[M(r,v)+vb_{0}(r)
+2va(r,v)\right]^{\frac{3}{2} } }
    = \frac{1}{2} \sqrt{v} \frac{\frac{dx}{dr} }{x^{\frac{3}{2} } }     =B(r,v) \; ,
\end{equation}
or
\begin{equation}\label{GrindEQ5}
  \frac{dx}{dr} =\frac{1}{\sqrt{v} } \left[2B(r,v)x^{\frac{3}{2} } \; ,
\right]\equiv f(x,r) \; ,
\end{equation}
   with the initial condition
\begin{equation} \label{GrindEQ6}
  x(0,v)=M_0(v)+vb_{0} (0)+2va(0,v) \; .
\end{equation}

We now  ensure the existence of a $\mathcal{C}^1$-function
$x(r, v)$ as a solution of the above initial value problem
defined throughout the cloud. The function $f(x, r)$ is
continuous in $r$, with $x$ restricted to a bounded domain. With
such domain of $r$ and $x$, $f(x, r)$ is also a
$\mathcal{C}^1$-function in $x$ which means $f(x, r)$ is
Lipschitz continuous in $x$. Therefore, the differential equation
\eqref{GrindEQ5} has a unique solution satisfying the initial
condition \eqref{GrindEQ6}, provided $f(x, r)$ satisfies
a certain condition given below.

Further, we can ensure that the solution will be defined
over the entire cloud, {\it  i.e.} for all $r \in [0,r_b]$, by
using the freedom in the choice of arbitrary function $B(r, v)$.
For this, we consider the domain $[0, r_b] \times [0, d]$ for
some finite $d$.

Let  us take $S=\sup \left|f(x,r)\right|$.
Then the differential equation \eqref{GrindEQ5} has a unique
solution defined over the entire cloud provided,
\begin{equation}\label{GrindEQ7}
    r_{b} \le \inf (r_{b} ,\frac{d}{S} )=\frac{d}{S} \; .
\end{equation}
This condition is to be satisfied according to usual existence
theorems to guarantee existence of a unique solution (see for
example \cite{gen1}).
Equation \eqref{GrindEQ7} implies  $S\le \frac{d}{r_{b} }$
{\it i.e.},
\begin{equation}\label{GrindEQ8}
    \mathop{\max }\limits_{0\le r\le r_{b,} ,0\le v\le d}
    \left|\frac{1}
{\sqrt{v} } \left[2B(r,v)x^{\frac{3}{2} } \right]\right|\le \frac{d}{r_{b} } \; .
\end{equation}
This, in turn, will be satisfied if
\begin{equation} \label{GrindEQ9}
0\le   \left|B(r,v)x^{\frac{3}{2} }\right|\le \frac{d\sqrt{v} }{2r_{b} } \; ,
\end{equation}
for all $r \in [0, r_b]$.

The collapsing cloud may start with $r_b $ small enough
so that the expression  $\frac{d\sqrt{v} }{2r_{b} } $ which is always
positive, satisfies the condition \eqref{GrindEQ9} with $x$
restricted to a bounded domain. We then have infinitely many
choices for the function $B(r, v)$, which is continuous and
satisfies conditions \eqref{GrindEQ2} and \eqref{GrindEQ9} for
each choice of $v$. For each such $B(r, v)$, there will be a
unique solution $x(r, v)$ of the differential equation
\eqref{GrindEQ5}, satisfying initial condition \eqref{GrindEQ6},
defined over the entire cloud, and in turn, there exists
a unique function $b_0(r)$ for each such choice of $B(r, v)$,
that is given by the expression
\begin{equation} \label{GrindEQ10}
b_{0} (r)=\frac{x(r,v)-M(r,v)-2va(r,v)}{v} \; .
\end{equation}

Thus, we have shown the following: For a given constant
$v$-surface and given initial data of mass function $F(t, r) =
r^3M(r, v)$ and $a(r, v)$  satisfying physically reasonable
conditions (expressed on $M$), there exists infinitely many
choices for the function $b(r)$ such that the condition
\eqref{GrindEQ1} is satisfied.
The condition continues to hold as $v \rightarrow 0$,
because of continuity. Hence, for all these configurations
the central singularity developed
in the collapse is a naked singularity.

The above analysis shows that the initial data satisfying
conditions \eqref{GrindEQ}, \eqref{GrindEQ2} and \eqref{GrindEQ9}
lead the collapse to a naked singularity. If we change the sign
in condition \eqref{GrindEQ2}, call it condition \eqref{GrindEQ2}$'$,
then above analysis apply and the initial data satisfying
conditions \eqref{GrindEQ}, \eqref{GrindEQ2}$'$ and \eqref{GrindEQ9}
lead the collapse to a black hole.
In the cases discussed above, in addition to above conditions,
energy conditions are also to be satisfied, and we have shown
above that this is always possible for matter models leading
to both possible outcomes.

Thus, from the above analysis, we get the following
conditions which should be satisfied by the initial data in
order that the end state of collapse is a naked singularity:

\begin{enumerate}
  \item  Energy conditions: $\rho \geq 0$ , and  $\rho  + p \geq 0$.
  \item  Reality condition given by equation \eqref{GrindEQ} above.
  \item  Condition on $B(r,v)$: $B(0,v)  < 0$ for naked singularity
and $B(0,v)> 0$ for black hole.
  \item $0\le $  $\left|B(r,v)x^{\frac{3}{2} } \right|\le
\frac{d\sqrt{v} }{2r_{b} } $.
  \end{enumerate}

For convenience, we denote the function
$$
C(r,v)=\frac{M(r,v)}{v} +b_{0} (r)+2a(r,v) \; .
$$
Then the reality
condition (2) becomes $C(0,v) > 0 $. Assuming this condition,
condition (3) will be satisfied if and only if
$\frac{M'(0)}{v} +b_{0} ^{{'}} (0)+2a'(0,v)<0$ for naked singularity,
and $>$ 0 for a black hole. Whenever $C(r,v)$ is an increasing
function of $r$ in the neighborhood of $0$, we get its derivative
positive, and so $B(0,v)> 0$, and end state will be a black hole.
On the other hand, if $C(r,v)$ is a decreasing function of $r$ in
the neighborhood of $0$, we get its derivative negative, and so
$B(0,v) <0$, and the endstate will be a naked singularity.

Regarding condition (4), using the expression $x = vC(r,v)$, it
becomes $C'(r,v)\le \frac{d}{vr_{b} } $  which will be satisfied
if $v$ and $r_b$  are sufficiently small. Thus validity of all
these conditions does not put any stringent restrictions on the
initial data.

The conclusion then is the following: If the initial data
consisting of the mass function $M(r,v)$ and function $a(r,v)$
satisfies the above conditions, then there is a large class of
velocity functions $b_0(r)$ such that end state of collapse
is either a black hole or a naked singularity, depending on
the nature of function $C(r,v)$ as explained above.

\subsection{Measure of the set of initial data leading to naked singularity}

We now show that the set of initial data $ \mathcal{G} = \{M(r,v),
a(r,v), b(r) \}$ satisfying the above conditions which lead the
collapse to a naked singularity, is an open subset of $X \times X
\times X$, where $X$ is an infinite dimensional Banach space
of all $\mathcal{C}^{1}$ or $\mathcal{C}^{2}$ real-valued functions defined on
$ \mathcal{D} = [0 ,  r_b ] \times  [0 ,  d ]$, endowed with
the norms
\begin{eqnarray}
  \parallel \textit{M}(r,v) \parallel_{1} &=& \sup_{{\mathcal{D}} } |\textit{M}  |
  + \sup_{{\mathcal{D}} } |\textit{M}_{,r}| + \sup_{{\mathcal{D}} } |\textit{M}_{,v}| \; ,    \\
  \parallel \textit{M}(r,v) \parallel_{2} &=& \parallel \textit{M}(r,v) \parallel_{1}
  + \sup_{{\mathcal{D}} } |\textit{M}_{,rr}  | + \sup_{{\mathcal{D}} } |\textit{M}_{,rv}|  + \sup_{{\mathcal{D}} }
|\textit{M}_{,vv} | \; .
\end{eqnarray}

These norms are equivalent to the standard $\mathcal{C}^{1}$
and $\mathcal{C}^{2}$ norms,
\begin{eqnarray}
\parallel \textit{M}(r,v) \parallel_{1} &=& \sup_{{\mathcal{D}} } (|\textit{M}|+ |\textit{M}_{,r}| + |\textit{M} _{,v}|) \; ,\\
 \parallel \textit{M}(r,v) \parallel_{2} &=& \sup_{{\mathcal{D}}} (\parallel \textit{M}(r,v) \parallel_{1} | + |\textit{M}_{,rr}|
+ |\textit{M} _{,rv}|+ |\textit{M}_{,vv}|) \; . \label{def38}
\end{eqnarray}

Let ${\mathcal{G}_1} = \{ M(r,v): M> 0, M \text{ is } \mathcal{C}^{1},  \
E_{1}> 0  \text{ and } E_{2} > 0 \text{ on } \mathcal{D} \}$ be a
subset of $X$, where $ E_1 = 3M + rM_{,r} + rv'M_{,v}$ and $E_2 =
3M + rM_{,r} -vM_{,v}$. Thus $E_1>0$ and $E_2>0$ are equivalent to
energy conditions.

We first show that $\mathcal{G}_1$ is an open subset of $X$.
For simplicity, we use the $\mathcal{C}^{1}$ norm, but a similar proof holds for
the $\mathcal{C}^{2}$ norm also. For $M \in \mathcal{G}_1$, let us
put $\delta = \min(M)$, $\gamma = \min( E_{1})$ , $\beta = \min(
E_{2} )$, $\lambda_{1} = \max(v)$, $\lambda_{2} = \max( v')$ and
for $r$ varying in $[0,r_{b}]$ and $ v \in [0,1]$, the functions
involved herein are all continuous functions defined on a compact
domain $\mathcal{D}$ and hence, their maxima and minima exist.
We define a positive real number
 \begin{equation}
  \mu = \frac{1}{2} \min\{\delta , \frac{\gamma}{9},
\frac{\gamma}{3 r_{b}}, \frac{\gamma}{3 r_{b} \lambda_{2}}, \frac{
\beta} {9 }, \frac{ \beta} {3 r_{b}}, \frac{ \beta} {3 \lambda_{1}}\}.
\end{equation}

Let $M_{1}(r,v)$ be  $\mathcal{C}^{1}$ in $\mathcal{D}$ with $\parallel M - M_{1}
\parallel_{1}  < \mu$.
Using above definition  we get $| M_{1} -M | <\mu$,
$| M_{1,r} - M_{,r} | < \mu$, $|M_{1,v} - M_{,v} |< \mu$  over $\mathcal{D}$.
Therefore, for choice of $\mu$, the respective inequalities are
\begin{eqnarray}\label{de38}
M_{1} &>& M - \frac{\delta}{2} > 0 , \\ \nonumber
3 |M_{1} - M | &<& \frac{\gamma}{6} , \\ \nonumber
 r |M_{1,r} - M_{,r} | &\leq& r_{b}|M_{1,r} - M_{,r} | < \frac{\gamma}{6} , \\ \nonumber
 r v' | M_{1,v} - M_{,v} |  &\leq& r_{b} \lambda_{2}| M_{1,v} - M_{,v} | < \frac{\gamma}{6}, \\ \nonumber
3  |M_{1} - M | &<& \frac{\beta}{6} , \\ \nonumber
 r  |M_{1,r} - M_{,r} |  &<& \frac{\beta}{6} , \\ \nonumber
 v | M_{1,v} - M_{,v} | &\leq& r_{b} | M_{1,v} - M_{,v} | < \frac{\beta}{6},
 \end{eqnarray}
that are satisfied on ${\mathcal{D}}$.
The  $2^{nd}$ , $3^{rd}$ and $4^{th} $ inequalities from above yield
\begin{eqnarray}
 3 |\textit{M}_{1} - \textit{M} | + r
|\textit{M}_{1,r} - \textit{M}_{,r} | + r v'|\textit{M}_{1,v} -
\textit{M}_{,v} | < \frac{\gamma}{2} < {\gamma}  \leq E_{1} \; .
\end{eqnarray}
Further, we can write $|[3 M_{1} + rM_{1,r}+ r v' M_{1,v}] - E_{1}| < E_{1} $
where $ E_{1} > 0 $ on $\mathcal{D}$.

Hence, $ 3 M_{1} + r M_{1,r}+ r v' M_{1,v} > 0 $ on
$\mathcal{D}$. Using similar analysis for last four inequalities
of equation \eqref{de38}, we obtain $ 3 M_{1} + rM_{1,r}
-vM_{1,v} > 0$ on $\mathcal{D}$.

Thus, $ M_{1}  > 0$,
$M_{1}$ is $\mathcal{C}^{1}$ , $ [3 M_{1} + r
M_{1,r}+ r v' M_{1,v}] > 0 $ and $ [3
M_{1} + r  M_{1,r} - v M_{1,v} ] > 0$ on
$ {\mathcal{D}} $ \ provided $v > 0$  throughout ${\mathcal{D}}$.
Therefore, $M_{1}(r,v)$ also lies in ${\mathcal{G}_1}$ \
and hence,\ ${\mathcal{G}_1}$ is an open
subset of $X$.

Using the similar argument we can show that the set $\mathcal{G}_2
= \{M(r,v): |B(r,v)| x^\frac{3}{2} < \frac{d\sqrt{v}}{2r_b}
\}$ is also an open subset of $X$. Thus the set of $M(r,v)$
satisfying above conditions forms an open subset of $X$, since
intersection of finite number of open sets is open. Similar
arguments show that the set of $a(r,v)$ and $b(r)$ satisfying these
conditions form separately open subsets of $X$. Hence using
definition of product topology we see that the set $\mathcal{G}$
defined above is an open subset of $X\times X \times X$. Thus
initial data leading the collapse to a naked singularity forms an
open subset of the Banach space of all possible initial data, and
therefore it is generic.

We now discuss measure theoretic properties of the open
set $\mathcal{G}$ consisting of the initial data leading the
collapse to a naked
singularity. By referring to the relevant literature about measures
on infinite dimensional separable Banach spaces, we argue that
this $\mathcal{G}$ has strictly positive measure in an appropriate
sense. For simplicity, we consider a single space $X$ and its open
subset $\mathcal{G}$. We ask the question: Does there exist a
measure on $X$ which takes positive value on $\mathcal{G}$? To
answer this question, we note that $X$ is an infinite dimensional
separable Banach space, and it is a consequence of Riesz lemma in
functional analysis that every open ball in $X$ contains an
infinite disjoint sequence of smaller open balls. So, if we want a
translation invariant measure on $X$\, then its value will be same
on each of these balls. Thus, if we demand that the surrounding
ball has finite measure, then each of these smaller balls will
have measure zero. Otherwise sum of their measures would be
infinite by countable additivity. In other words, for separable
Banach spaces, every open set has either measure zero or infinite
under a translation invariant measure. So, if we wish to have a
non-trivial measure on $X$, then we have to discard the property
of translation invariance. In that case we must shift to Gaussian
measures or Wiener measures. Under these measures, we can conclude
that an open subset of $X$ will have a positive measure.
For example, it is proved in \cite{gen2} (Theorem 2 on page 159),
that Gaussian measure of
an open ball in a separable Banach space is positive. We can also
use Wiener measure on $\mathcal{C}([0,1])$to get the same result (see for
example \cite{gen3} and \cite{gen4}).

However, for all practical purposes in Physics and
Astrophysics, physical functions could be assumed to be Taylor
expandable. Thus assuming that our initial data is regular and
Taylor expandable, and again working for simplicity with a single
function space, instead of product space, we can formulate our
problem of measure as follows: Let $Y$ denote the space of Taylor
expandable functions defined on an interval $[0,T]$. We consider
initial data consisting of functions with first finite number of
terms, say $n$ terms, which lead the collapse to a naked
singularity. These functions will belong to a finite dimensional
space isomorphic to $\mathbb{R}^n$. Working with supremum
norm as above and
arguing similarly, we can prove that the initial data set
satisfying conditions (1) to (4) above is an open subset of $\mathbb{R}^n$.
Now, we have a standard result (see for example \cite{gen5},
prop. 4.3.4, page 83) that a Lebesgue
measure of an open subset in $\mathbb{R}^n$ is strictly positive.
Denoting this measure by $\mu_{n}$ and the open set by
$\mathcal{G}_{n}$, we get $ \mu_{n}(\mathcal{G}_{n}) > 0 $.
If, further, $\mathcal{G}_{n}$ is bounded, then
$ \mu_{n} (\mathcal{G}_{n})$ will be finite. Hence normalized Lebesgue
measure of an open subset, and in particular, of an open ball in
$\mathbb{R}^n$ is also strictly positive, and in fact bounded. We ask the
question: Assuming that $ \mu_{n} (\mathcal{G}_{n})> 0$,
can we get a measure on $Y$ such that measure of
$\mathcal{G}$ is positive? This is answered affirmatively by
Maxwell-PoincarÃ¨ theorem which is stated as follows (see for
example \cite{gen6}):\\
Consider the sequence of the normalized Lebesgue measures on the Euclidean
spheres $ S_{r_{n}}^{n-1} \subset {\mathbb R}^{n}  $ of radius
$r_{n} = c \sqrt{n} , c >0 $ and the limit of spaces
$$
{\mathbb R}^{1} \subset {\mathbb R}^{2} ... \subset
{\mathbb R}^{n} \subset ...  \subset {\mathbb R}^{\infty} \; .
$$
Then the weak limit of these measures is the standard Gaussian
measure $\mu$ which is the infinite product of the identical Gaussian
measures on the line with zero mean and variance $c^2$. Thus the
limit exists and is positive on an open subset $\mathcal{G}$.

It is also possible to give other approaches which
answers affirmatively the existence of such limits which are
termed as infinite products or in general `projective limits'.
We describe briefly one such approach as described by
Yamasaki \cite{gen7} (For general concepts on measure theory,
we refer to \cite{gen8}). \\
Let $\mathbb{R}^{\infty} $ denote the infinite product
of real lines. Let $\mathbb{R}_{0}^{\infty}$ be the subspace
of $\mathbb{R}^{\infty}$ given by
$ \mathbb{R}_{0}^{\infty} =\{(\xi_{n})$: there exists $ N$ with
$\xi_{n}= 0$ for $n \geq N \} $. Then $\mathbb{R}^{\infty}$ is the
algebraic dual of
$\mathbb{R}_{0}^{\infty}$ and  ${\mathcal{B}}_{  \mathbb{R}_{0}^{\infty}}$
is the weak Borel field of  $\mathbb{R}^{\infty}$.
Members of ${\mathcal{B}}_  {\mathbb{R}_{0}^{\infty}}$ are called weak
Borel subsets of $\mathbb{R}^{\infty}$. The space $Y$ mentioned
above can be seen isomorphic to a subspace of $\mathbb{R}^{\infty} $,
and is isomorphic to $\mathbb{R}_{0}^{\infty}$ if we consider a finite
number of terms in the Taylor expansion. Let $m$ denote one
dimensional Lebesgue measure on $\mathbb{R}$. Let $G = (G_{k})$ be a sequence
of Borel sets of $\mathbb{R}$ such that $0 < m(G_{k}) < \infty $.
We shall define two Borel measures $m_{k}$ and $\lambda_{k}$ by
$$
m_{k}(B) = m(B) / m(G_{k}), \; \lambda_{k}(B) = m_{k}(B \cap G_{k}),
$$
$m_{k}$ is $\sigma$-finite, whereas $\lambda_{k}$ is a
probability measure on $ \mathbb{R}$.

Consider the product measure
$$
\mu_{G}^{(n)} = \prod_{k=1}^{n} m_{k} \times \prod_{k=n+1}^{\infty} \lambda_{k} \; ,
$$
then $\mu_{G}^{(n)}$ is $\sigma$-finite on $\mathbb{R}^{\infty}$. \\
Then we have the following theorem:
For every weak Borel set $E$ of $\mathbb{R}^{\infty}$, put
$$
\mu_{G}^{(E)} = \lim_{n \rightarrow \infty} \mu_{G}^{(n)} (E).
$$
This limit always exists and becomes a $\sigma$-finite
$\mathbb{R}_{0}^{\infty} $-invariant measure on
$\mathbb{R}^{\infty}$. Then $\mu_{G}$ lies on
$$
L(G) = \cup_{n = 1}^{\infty} L_{n}(G),
$$
where
$$
L_{n}(G) = \mathbb{R}^{n} \times \prod_{k = n+1}^{\infty} G_{k} \; .
$$
We note that the measure $\mu_{G}$ defined in this theorem is
called the infinite dimensional Lebesgue measure supported
by $L(G)$.

For any open bounded subset $E$ of $\mathbb{R}^{\infty}$, $E$ is a Borel set and hence a weak Borel set. Moreover
\begin{equation}\nonumber
 \mu_{G}^{(n)}(E) = \left( \prod_{k=1}^{n} m_{k} \times \prod_{k=n+1}^{\infty} \lambda_{k} \right) (E)
= \left( \prod_{k=1}^{n} m_{k}\right) (E) \times \left( \prod_{k=n+1}^{\infty} \lambda_{k} \right) (E) \; ,
\end{equation}
and both these factors are finite. Thus the measure $\mu_{G}$ takes a non-zero value on an open bounded subset $E$ of $\mathbb{R}^{\infty}$.
We can employ this measure instead of the Gaussian measure mentioned in Maxwell-PoincarÃ¨ theorem to yield the desired result. In any case, use of probability measure is inevitable and we conclude that the space of initial data leading to a certain outcome (be it black hole or naked singularity), within a specific collapse scenario has non zero measure with respect to the set of all possible initial data.

\section{Equation of state}\label{eos}
As is known the presence of an equation of state
introduces a differential relation for the previously considered
free function that closes the system of Einstein equations.
Examples of simple, astrophysically relevant, linear and polytropic
equations of states are discussed below.

In the scenario described above, the relation between
the density and pressure could vary during collapse,
as it is natural to assume in the case where we go from a
nearly Newtonian initial state to a final state where a very
strong gravitational field is present. The equation relating
$p$ to $\rho$ will therefore be represented by some function
of $r$ and $v$ that is related to the choice of the free function
$M$. There are at present many indications that suggest how
in the presence of high gravitational fields gravity can act
repulsively and pressures can turn negative towards the end of
collapse. Therefore if that is the case then the equation of
state relating pressure and density (which is always positive)
must evolve in a non-trivial manner during collapse.

Typically we can expect an adiabatic behaviour with small
adiabatic index at the beginning of collapse when the energy
density is lower than the nuclear saturation energy. It is
not unrealistic to suppose that the equation of state will
have sharp transitions when matter passes from one regime
to another, as is the case when the limit of the nuclear saturation
energy is exceeded. Towards the end of collapse repulsive
forces become relevant thus giving rise to negative pressures
and the speed of sound approaches the speed of light
\cite{Zeld}.

Nevertheless it is interesting to analyze the structure
of collapse model within one specific regime once a fixed
equation of state, of astrophysical relevance, is imposed. If we
choose the equation of state to be linear barotropic or polytropic
we can describe collapse of the star right after it departs
from the equilibrium configuration where gravity was balanced
by the nuclear reactions taking places at its center.
From an astrophysical point of view, neglecting the energy coming
from the nuclear reactions occurring at the interior is reasonable
since we know that once the nuclear fuel of the star is exhausted
the star is subject to its own gravity only and the departure
from equilibrium occurs in a very short time.
In this sense equilibrium models for stars (such as the early
models studied in the pioneering work by Chandrasekhar
\cite{Chandra})
constitute the initial configuration of our collapse model
and the physical parameters used to construct those equilibrium
models will translate into the initial conditions for density
and pressure.

As we mentioned before, introducing an equation of state
is enough to ensure that the system of Einstein equations is
closed and so no freedom to specify any function remains. In fact
a barotropic equation of state of the form,
\begin{equation}
    p=Y(\rho) \; ,
\end{equation}
introduces a differential equation that must be satisfied
by the mass function $M(r,v)$, thus providing the connection
between equation \eqref{p} and equation \eqref{rho} and making
them dependent on $R(r,v)$ and its derivatives only.  The dynamics
is entirely determined by the initial configuration
and therefore we see how solving the equation of motion \eqref{Rdot}
is enough to solve the whole system of equations.

In this case solving the differential equation for $M$ might
prove to be too complicated. Nevertheless with the assumption that
$M$ can be expanded in a power series as in equation \eqref{powerM}
we can obtain a series of differential equations for each order $M_i$.
Expanding the pressure and density near the center we obtain
explicitly the differential equations that, if they can be satisfied
by all $M_i$ converging to a finite mass function $M$, solve the
problem, thus giving the explicit form of $M$.

From
\begin{eqnarray}\nonumber
  p(r,v) &=& p_0(v)+p_2(v)r^2+...= Y_0(v)+Y_2(v)r^2+... \; , \\
  \rho(r,v) &=& \rho_0(v)+\rho_2(v)r^2+... \; ,
\end{eqnarray}
with
\begin{equation}
  Y_0 = Y(\rho_0)  \; , \;  Y_2(v) = Y_{,\rho}\rho_2 \; ,
\end{equation}
and from Einstein equations \eqref{p} and \eqref{rho} we get
\begin{eqnarray}
  Y(\rho_0) &=& -\frac{M_{0,v}}{v^2} \; ,\\
  Y_{,\rho}\rho_2 &=& -\frac{M_{2,v}}{v^2} \; ,
\end{eqnarray}
with
\begin{eqnarray}
  \rho_0 &=& \frac{3M_0}{v^3} \; ,\\
  \rho_2 &=& \frac{5M_2}{v^3}-\frac{3M_0}{v^4}w_{,r}(0,v) \; .
\end{eqnarray}
Once again we see that without the knowledge of $w$,
which is related to $v'$, it is impossible to solve the set of
differential equations in full generality. Furthermore we can see
that whenever pressures and density can be expanded in a power
series near the center the behaviour close to $r=0$ approaches
that of an homogeneous perfect fluid.

There are a few equations of state that have been widely
studied in equilibrium models for stars and that naturally translate
into collapse models.
The simplest one is a linear equation of state of the form
\begin{equation}\label{linear}
    p=\lambda\rho \; ,
\end{equation}
where $\lambda$ is a constant. This case was studied in
\cite{Joshi-Goswami}
where it was shown the existence of a solution of
the differential equation for $M$, which, from Einstein equations
\eqref{p} and \eqref{rho} becomes
\begin{equation}
    3\lambda M+\lambda r M_{,r}+[v+(\lambda+1)rv']M_{,v}=0 \; .
\end{equation}
It was shown that both black holes and naked singularities
are possible outcomes of collapse depending on the initial data
and the velocity distribution of the particles.

Another possibility is given by a polytropic equation
of state of the type
\begin{equation}\label{polytropic}
    p=\lambda\rho^\gamma \; .
\end{equation}
Such an equation of state is often used in models for stars
at equilibrium and can describe the relation between $p$ and $\rho$
in the early stages of collapse.
Therefore the physical values for $p_0$, $\rho_0$, $\lambda$
and $\gamma$ at the initial time can be taken from such models at
equilibrium and expressed in terms of the thermodynamical
quantities of the system such as the temperature and the molecular
weight of the gas. The pressure is typically divided in a matter
part (describing an ideal gas) and a radiation part (related to
the temperature). The exponent $\gamma$ is generally written
as $\gamma=1+1/n$, where $n$ is called polytropic index of the
system and is constrained by $n\leq 5$ (for $n>5$ the cloud has
no boundary at equilibrium)
\cite{Tooper1}, \cite{Tooper2}.
The formalism developed above
can therefore be used to investigate such realistic scenarios
for collapse of massive stars.

\section{Concluding remarks}\label{remarks}

We have studied here the general structure of complete
gravitational collapse of a sphere composed of perfect fluid
without a priori requiring an equation of state for the matter
constituents, thus allowing for the freedom to choose the mass
function arbitrarily, as long as physical reasonableness as
imposed by regularity and energy conditions is satisfied.

The interest of such an analysis comes from the fact
that the class of perfect fluid models for matter is considered
to be physically viable for the description of realistic objects
in nature such as massive stars and their gravitational collapse.
Typically perfect fluids are considered to be physically more
sound than models where matter is approximated by dust-like behaviour
{\it i.e.} without pressures, or where matter is sustained by only
tangential pressures (though the `Einstein cluster' describing
a spherical cloud of counter-rotating particles has been shown to
have some non-trivial  physical validity
\cite{cluster1}-\cite{cluster3}).
What we have shown is that, within the class of perfect
fluid collapses, both final outcomes, namely black holes and
naked singularities, can be equally possible depending on
the choice of the initial data and the free function $F$.
In fact our results show that naked singularities and black
holes are both possible final states of collapse, much in the same
way as it has already been proven in the simpler cases of
inhomogeneous dust and matter exhibiting only tangential stresses.
The sets of initial data leading to either of the outcomes share
the same properties in terms of genericity and stability.

The structure of initial data sets in the case
of OSD, LTB and pressure collapse and their inter-relationship
is, however, a complicated issue. Nevertheless, we can
comment on this based on the studies in
\cite{Sar-Pramana}, \cite{SG},
and the results proved in sections \ref{perturbation} and
\ref{genericity} in this paper. In \cite{SG} it was proven that
the space of initial data $\{ M(r), b( r )\}$ leading LTB collapse
to black hole or naked singularity forms an open subset of
$X \times X$, where $X$ is the infinite dimensional Banach space
of real $\mathcal{C}^1$ functions defined on the domain. As per the
analogous results in the case of non-vanishing tangential pressures
it follows that the initial data set leading
the collapse to OSD black holes is a non-generic subset of space
of $\{  M(r), b( r )\}$. The shortcoming of the tangential pressure
case being that is not wholly physically satisfactory. Therefore
to investigate the perfect fluid case, we moved to a `bigger'
space $X \times X \times X$, since the initial data set comprises
of $\{M(r, v_i), p (r, v_i), b( r)\}$. Thus, in this space,
the initial data set $I_{OSD}$ or $I_{TBL}$ or the union of both the
sets, will become non-generic. Mathematically speaking, this set
is meager or nowhere dense in the space $X \times X \times X$. This
is proved by the study performed here. Thus, the initial
configurations for the end states in the case of LTB or OSD models
lie on the critical surface separating the two possible outcomes
of collapse as discussed above.

This analysis in fact shows how the structure of Einstein
equations is very rich and complex, and how the introduction of
pressures in the collapsing cloud opens up a lot of new possibilities
that, while showing many interesting dynamical behaviours, do not
rule out either of the two possible final outcomes.

There are physical reasons, however, for the perfect fluid
model to be subdued to the choice of an equation of state and
there is also increasing evidence that such an equation of state
cannot hold during the whole duration of the dynamical collapse.
In fact there are indications that as the collapsing matter
approaches the singularity large negative pressures arise, thus making
the equation of state relating density to pressures depart from
usual well-known equations of stellar equilibrium. Nevertheless the
study of similar scenarios with linear or polytropic equations of
state can give insights in the initial stages of collapse of a star.
All this is very important from astrophysical point of view where
still little is known of the processes that happen towards the very
end of the life of a star, when in a catastrophic supernova explosion
the outer layers are expelled and the inner core collapses under
its own gravity.

As we mentioned, due to the intrinsic complexity of Einstein
equations for perfect fluid collapse, it is generally possible to
solve the system of equations only under some simplifying assumptions
(like the choice of a specific mass function), and only close to
the center of the cloud. The indications provided by the present
analysis are then a first step towards a better understanding of
what happens in the last stages of the complete gravitational
collapse of a realistic massive body.

Furthermore the above formalism could possibly be used as
the framework upon which to develop possible numerical simulations
of gravitational collapse. As seen in the comoving frame, the
positivity of $\chi_1$ (or $\chi_2$)) is the necessary and
sufficient condition for
the singularity to be visible, at least locally. Numerical models
of a collapsing star made of a perfect fluid with a polytropic
equation of state (or a varying equation of state that takes into
account the phase transitions that occur in matter under strong
gravitational fields), with the addition of rotation and possibly
electromagnetic field might help us better understand whether the
inner ultradense region that forms at the center of the collapsing
cloud when the apparent horizon is delayed, might be visible globally
and have some effects on the outside universe. Many numerical
models that describe dynamical evolutions leading to the formation
of black holes exist both in gravitational collapse as in merger
of compact objects such as neutron stars
(see e.g. \cite{Rezzolla1}, \cite{Rezzolla2}).
but a fully comprehensive picture of what happens in the
final moments of the life of a star is still far away.

Close to the formation of the singularity, gravitationally
repulsive effects, possibly due to some quantum gravitational
corrections, are likely to take place. If such phenomena
can interact with the outer layers of the collapsing cloud they
might create a window to the physics of high gravitational fields
whose effects might be visible to faraway observers. This scenario
might in turn imply the visibility of the Planck scale physics or
new physics close to the singularity, the presence of a quantum wall
that might cause shock-waves from within the Schwarzschild radius
that might give rise to different type of emissions and explosions,
with photons or high energy particles escaping from the ultradense region.
Collisions of particles with arbitrarily high center of mass energy
near the Cauchy horizon might also happen
\cite{accel}.

Overall, the analysis led over the past few years seems
to suggest that the Oppenheimer-Snyder-Datt scenario is indeed
too restrictive to account for the richness of realistic dynamical
models in general relativity. The occurrence of naked singularities
in gravitational collapse appears to be a well established fact
and a lot of intriguing new physics might arise from the
future study of more detailed collapse models.

\end{document}